\documentclass[12pt]{article} 
\usepackage[utf8x]{inputenc} 
\usepackage[T1]{fontenc} 
\usepackage{amsmath,amssymb,bbm,commath} 
\usepackage{slashed} 
\usepackage[unicode]{hyperref} 
\usepackage{graphicx} 
\usepackage{enumerate} 
\hypersetup{bookmarksnumbered=true, bookmarksopen=true, bookmarksopenlevel=2, breaklinks=true, citecolor=blue, colorlinks=true, linkcolor=red, linktoc=page, pdfborder={0 0 0}, pdfkeywords={Semichirals, }{N=(2,2) Multiplets, }{Localization}, pdfstartview=FitH, pdfauthor={FBenini, PMCrichigno, DJain}, pdfsubject={Semichiral Fields on Sphere}, pdftitle={Semichiral Fields on S² and Generalized Kähler Geometry}, plainpages=false, unicode=true, urlcolor=cyan}
\usepackage{cite} 

\usepackage{geometry} 
\usepackage{fancyhdr} 
\usepackage{parskip} 
\usepackage[nottoc,notlof,notlot]{tocbibind} 
\usepackage[titles]{tocloft} 

\DeclareUnicodeCharacter{"0393}{\Gamma}
\DeclareUnicodeCharacter{"0394}{\Delta}
\DeclareUnicodeCharacter{"0398}{\Theta}
\DeclareUnicodeCharacter{"039B}{\Lambda}
\DeclareUnicodeCharacter{"039E}{\Xi}
\DeclareUnicodeCharacter{"03A0}{\Pi}
\DeclareUnicodeCharacter{"03A3}{\Sigma}
\DeclareUnicodeCharacter{"03A5}{\Upsilon}
\DeclareUnicodeCharacter{"03A6}{\Phi}
\DeclareUnicodeCharacter{"03A8}{\Psi}
\DeclareUnicodeCharacter{"03A9}{\Omega}
\DeclareUnicodeCharacter{"03B1}{\alpha}
\DeclareUnicodeCharacter{"03B2}{\beta}
\DeclareUnicodeCharacter{"03B3}{\gamma}
\DeclareUnicodeCharacter{"03B4}{\delta}
\DeclareUnicodeCharacter{"03B5}{\epsilon}
\DeclareUnicodeCharacter{"03B6}{\zeta}
\DeclareUnicodeCharacter{"03B7}{\eta}
\DeclareUnicodeCharacter{"03B8}{\theta}
\DeclareUnicodeCharacter{"03D1}{\vartheta}
\DeclareUnicodeCharacter{"03B9}{\iota}
\DeclareUnicodeCharacter{"03BA}{\kappa}
\DeclareUnicodeCharacter{"03BB}{\lambda}
\DeclareUnicodeCharacter{"03BC}{\mu}
\DeclareUnicodeCharacter{"03BD}{\nu}
\DeclareUnicodeCharacter{"03BE}{\xi}
\DeclareUnicodeCharacter{"03C0}{\pi}
\DeclareUnicodeCharacter{"03C1}{\rho}
\DeclareUnicodeCharacter{"03C3}{\sigma} 
\DeclareUnicodeCharacter{"03C4}{\tau}
\DeclareUnicodeCharacter{"03C5}{\upsilon}
\DeclareUnicodeCharacter{"03C6}{\phi}
\DeclareUnicodeCharacter{"03D5}{\varphi}
\DeclareUnicodeCharacter{"03C7}{\chi}
\DeclareUnicodeCharacter{"03C8}{\psi}
\DeclareUnicodeCharacter{"03C9}{\omega}
\DeclareUnicodeCharacter{"21D0}{\Leftarrow}
\DeclareUnicodeCharacter{"0212B}{\AA}
\DeclareUnicodeCharacter{"00B7}{\cdot}
\DeclareUnicodeCharacter{"266A}{\eighthnote}
\DeclareUnicodeCharacter{"266B}{\twonotes}
\PrerenderUnicode{²}\PrerenderUnicode{ä}

\newcommand{\TPheader}[3] {\thispagestyle{fancy}\pagenumbering{alph}\lhead{#1}\chead{#2}\rhead{#3}\cfoot{}}
\newcommand{\makepage}[1] {\newpage\pagenumbering{#1}}

\newcommand{\Abstract}[1] {\begin{abstract}\small #1 \end{abstract}}

\newcommand{\bea}{\begin{equation}\begin{aligned}}
\newcommand{\eea}{\end{aligned}\end{equation}}

\newcommand\eqs[1] {\begin{align}#1\end{align}}
\newcommand\eqsn[1] {\begin{align*}#1\end{align*}}
\newcommand\eqsc[1] {\begin{gather}#1\end{gather}}

\newcommand\equ[1] {\begin{equation}#1\end{equation}}
\newcommand\equn[1] {\begin{equation*}#1\end{equation*}}

\newcommand\pmat[1] {\begin{pmatrix}#1\end{pmatrix}}

\renewcommand\( {\left(}
\renewcommand\) {\right)}

\newcommand\D {{\cal D}}

\renewcommand\L {{\cal L}}
\newcommand\M {{\cal M}}
\newcommand\N {{\cal N}}

\newcommand\Q {{\cal Q}}

\renewcommand\Bbb {\mathbb}

\newcommand\Bd {{\Bbb D}}

\newcommand\CM{{\mathcal M}}

\numberwithin{equation}{section} 
\begin{document}
\title{\LARGE \bf On gauged linear sigma models with torsion}

\author{P. Marcos Crichigno$^{a}$\footnote{\href{mailto:p.m.crichigno@uu.nl}{p.m.crichigno@uu.nl}}\,  and Martin Ro\v{c}ek$^{b}$\footnote{\href{mailto:martin.rocek@stonybrook.edu}{martin.rocek@stonybrook.edu}} 
\bigskip \\
\normalsize 
\emph{$^{a}$Institute for Theoretical Physics and}\\
\normalsize \emph{Center for Extreme Matter and Emergent Phenomena}\\
\normalsize \emph{Utrecht University, Utrecht 3854 CE, The Netherlands}
\bigskip\\ 
\normalsize \emph{$^{b}$C. N. Yang Institute for Theoretical Physics}\\
\normalsize \emph{State University of New York, Stony Brook, NY 11790-3840} }

\date{} 

\maketitle

\TPheader{\today}{}{ITP-UU-15/09\\YITP-SB-15-14}

\Abstract{
\normalsize

We study a broad class of two dimensional gauged linear sigma models (GLSMs) with off-shell $\N=(2,2)$ supersymmetry that flow to nonlinear sigma models (NLSMs) on noncompact geometries with torsion. These models arise from coupling chiral, twisted chiral, and semichiral multiplets to known as well as to a new $\N=(2,2)$ vector multiplet, the constrained semichiral vector multiplet (CSVM).  We discuss three kinds of models, corresponding to torsionful deformations of standard GLSMs realizing K\"ahler, hyperk\"ahler, and Calabi-Yau manifolds. The $(2,2)$ supersymmetry guarantees that these spaces are generalized K\"{a}hler. Our analysis of the geometric structure is performed at the classical level, but we also discuss quantum aspects such as R-symmetry anomalies. We  provide an explicit example of a generalized K\"{a}hler structure on the conifold. 
}

\makepage{Roman} 
\tableofcontents
\makepage{arabic}

\section{Introduction} 
\label{Introduction}

Two-dimensional nonlinear sigma models (NLSMs) with $\N=(2,2)$ supersymmetry are an essential tool in string theory. When the $b$-field is a closed two-form, $H=db=0$, they describe strings propagating on K\"{a}hler backgrounds. For $H$ not necessarily zero, they describe strings propagating on generalized K\"{a}hler manifolds, the most general target space of $\N=(2,2)$ NLSMs \cite{Gates:1984nk}. Due to their nonlinear character, the nonperturbative quantum properties of NLSMs can be difficult to untangle, even with a large amount of supersymmetry. An approach that has proven to be extremely successful is to realize  NLSMs as gauged linear sigma models (GLSMs) in the UV \cite{Witten:1993yc}. Although much is understood about the gauge theory description of NLSMs on K\"{a}hler manifolds, much less is known about generalized K\"{a}hler manifolds.

In this paper we describe certain GLSMs with off-shell $\N=(2,2)$ supersymmetry realizing NLSMs on noncompact generalized K\"{a}hler manifolds. These models arise from coupling semichiral multiplets to known as well as to new $\N=(2,2)$ vector multiplets which we describe here.  The generalized K\"{a}hler structure is controlled by a set of continuous parameters, and only for a special value of such parameters it becomes K\"{a}hler. 

Previous work on GLSMs for semichiral fields includes \cite{Merrell:2006py,Adams:2006kb}; more general couplings of gauge fields to semichiral fields were discussed in \cite{Lindstrom:2007vc,Merrell:2007sr,Lindstrom:2007sq,Lindstrom:2008hx,Crichigno:2011aa}. The gauging of sigma models with a Wess-Zumino term was studied in \cite{Hull:1989jk,Jack:1989ne}. The case with on-shell $\N=(2,2)$ supersymmetry was studied in \cite{Kapustin:2006ic}. For $\N=(0,2)$ GLSMs with torsion see \cite{Adams:2006kb,Adams:2009av,Quigley:2011pv,Blaszczyk:2011ib,Adams:2012sh,Quigley:2012gq,Melnikov:2012nm}.

Let us briefly review some relevant aspects of generalized K\"ahler geometry. A generalized K\"{a}hler structure on a manifold $\mathcal M$ consists of the triplet $(g,J_{\pm},H)$, where $g$ is a Riemannian metric, $J_{\pm}$ are two integrable complex structures, and $H$ is a closed three-form (which locally can be written as $H=db$). The complex structures are covariantly conserved, $i.e.$, $\nabla^{\pm} J_{\pm}=0$ with respect to a connection with torsion $\nabla^\pm=\nabla^0\pm \tfrac{1}{2} g^{-1}H$, where $\nabla^0$ is the Levi-Civita connection. The presence of torsion implies that the geometry is generically not K\"{a}hler: the forms $\omega_{\pm}=g J_{\pm}$ are not closed. This structure was originally discovered in the context of nonlinear sigma models in \cite{Gates:1984nk}, where it was called bihermitian geometry. More recently, it was reformulated as the analogue of K\"{a}hler geometry in the
context of generalized complex geometry \cite{Gualtieri:2003dx,2010arXiv1007.3485G}. For introductory lectures on generalized complex geometry and its relation to supersymmetry, see for instance \cite{Zabzine:2006uz,Koerber:2010bx}. For a review of generalized K\"{a}hler geometry and general $\N=(2,2)$ NLSMs see \cite{Lindstrom:2005zr}.

As shown in  \cite{Ivanov:1994ec,Lyakhovich:2002kc,Lindstrom:2005zr}, one may always choose (locally) coordinates in the generalized K\"{a}hler manifold $\CM$ which are adapted to the decomposition of $T^*\CM = \text{ker} (J_+-J_-)\oplus\text{ker} (J_++J_-)\oplus\text{coim} [J_+,J_-]$. In terms of the $\N=(2,2)$ sigma model, each of these directions is parametrized by different multiplets: chiral ($\Phi$), twisted chiral ($\chi$), left and right semichiral ($\Bbb X_L,\Bbb X_R$), respectively. As in K\"{a}hler geometry, the geometric data is locally encoded by a single scalar function $K$, the generalized K\"{a}hler potential, which serves as the action of the NLSM in $\N=(2,2)$ superspace \cite{Buscher:1987uw}:
\equ{\label{generic NLSM intro}
\L=\int d^4\theta \,K(\Phi,\bar \Phi;\chi,\bar \chi; \Bbb X_L,\bar{ \Bbb X}_L,\Bbb X_R, \bar{ \Bbb X}_R )\,,
}
where $d^4 \theta = d \theta^+ d\theta^- d \bar \theta^+ d\bar \theta^-$ is the usual Grassmann measure. Apart from certain inequalities that guarantee the target metric is positive definite, $K$ is arbitrary. If $K$ depends only on chiral (or only twisted chiral) fields, the target geometry is necessarily K\"{a}hler. If it depends on semichiral fields, or both chiral and twisted chiral fields the target manifold $\CM$ typically has torsion. The GLSMs we discuss here are described, at low energies, by a NLSM with an action of the form (\ref{generic NLSM intro}) which generically contains all three kinds of multiplets. As we shall discuss, the number of each kind of multiplet depends on the specific model. 

One may distinguish different classes of GLSMs depending on how the gauge symmetry acts and correspondingly, which vector multiplet couples to the model: ($i$) chiral(twisted chiral) and semichiral fields can be coupled to the standard(twisted) vector multiplets, respectively; ($ii$) semichiral fields can be coupled to the semichiral vector multiplet (SVM); and ($iii$) semichiral fields can be coupled to a new kind of vector multiplet, the constrained semichiral vector multiplet (CSVM).\footnote{One can also consider models involving gauge symmetries that act on chiral and twisted chiral fields simultaneously; these involve the Large Vector Multiplet \cite{Lindstrom:2007vc,Lindstrom:2008hx,Ryb:2007pf} and are not discussed here.} As we shall see, in all these cases, the gauge theories are continuous deformations of standard gauge theories realizing NLSMs on noncompact K\"{a}hler manifolds. The deformations are controlled by a set of continuous parameters $\beta_i$ which determine (among other couplings) a gauged $b$-field term in the action. The deformations typically break rigid flavor symmetries but preserve R-symmetry at the quantum level. We study the classical moduli space of these theories and show that they lead to a family of generalized K\"{a}hler structures on the Higgs branch. Class ($i$) leads generically to deformations of K\"{a}hler manifolds, class ($ii$) leads to deformations of hyperk\"{a}hler manifolds, and class ($iii$) leads to deformations of GLSMs realizing Calabi-Yau manifolds. Our analysis of the target space geometry is carried out only in the UV---an important question is what is the behavior of these theories in the deep IR, but we do not address this here.

The GLSMs studied here provide a useful method to study explicit generalized K\"{a}hler metrics.  In Section~\ref{Conifold with torsion} we discuss a particular example in class ($iii$) which describes a family of generalized K\"{a}hler structures on the resolved conifold. Although our main focus in this paper are geometries with torsion, the generalized K\"{a}hler description of a manifold is useful even in the absence of torsion. For instance, any hyperk\"{a}hler manifold admits a description in terms of semichiral fields \cite{Lindstrom:2005zr} and the generalized potential gives not only the metric and (closed) $b$-field, but also all three anticommuting complex structures. See \cite{Crichigno:2011aa,Dyckmanns:2011ts} for the semichiral description of gravitational instantons.

In recent years, the application of supersymmetric localization techniques \cite{Witten:1988xj,Witten:1991zz,Pestun:2007rz} to the two dimensional case \cite{Benini:2012ui,Doroud:2012xw} has shed new light into standard $\N=(2,2)$ GLSMs and their relation to various aspects of K\"{a}hler geometry, $e.g.$ the works \cite{Jockers:2012dk,Gomis:2012wy,Park:2012nn,Doroud:2013pka}. The supersymmetric localization on $S^2$ of the GLSMs described here is studied in \cite{Benini:2015isa}. 

\medskip

This paper is organized as follows. In Section~\ref{Background} we review some background material. In Section~\ref{GLSMs for Semichiral Fields} we introduce the constrained semichiral vector multiplet and its coupling to semichiral matter. In Section~\ref{Low energy behavior} we discuss the low energy behavior of various GLSMs for semichiral fields and the generalized K\"{a}hler structure on the moduli space. In Section~\ref{Conifold with torsion} we give an explicit example; the conifold with torsion and finally in Section~\ref{Discussion} we discuss open questions and future work.

\section{Background}
\label{Background}

\subsection{$\N=(2,2)$ SUSY and matter multiplets}

We begin by reviewing some basic elements of 2d $\N=(2,2)$ supersymmetry in flat space with Lorentzian signature. The algebra of $\N=(2,2)$ spinor derivatives is
\equ{
\{ \Bbb D_{\pm}, \bar{\Bbb D}_{\pm} \} = 2  i \, \partial_{\pm\pm}\,,
}
with all other anticommutators vanishing. These anticommute with the supersymmetry generators $\Bbb Q_\pm,\bar{\Bbb Q}_\pm$. 

A chiral multiplet $\Phi$ and a twisted chiral multiplet $\chi$ are defined by the supersymmetric constraints  
\equ{\bar \Bd_{+}\Phi=\bar \Bd_{-}\Phi=0\,, \qquad  \bar \Bd_{+}\chi=\Bd_{-}\chi=0\,,}
as well as their dual complex linear and twisted linear superfields \cite{Gates:1980az,Gates:1983nr} ($\Xi$ and $\tilde\Xi$, respectively)
\equ{\bar\Bd_+\bar\Bd_-\Xi=\Bd_+\Bd_-\bar\Xi=0\,, \qquad
\bar\Bd_+\Bd_-\tilde\Xi=\Bd_+\bar\Bd_-\bar{\tilde\Xi}=0\,.}
These multiplets are  well known and have been thoroughly studied; semichiral multiplets \cite{Buscher:1987uw}, however, less so. They come in two types, left and right, and are defined by the constraints
\equ{
\bar \Bd_{+} \Bbb{X}_{L}=\Bd_{+} \bar{\Bbb{X}}_{L}=0\,, \qquad \qquad \bar \Bd_{-} \Bbb{X}_{R}=\Bd_{-}\bar{ \Bbb{X}}_{R}=0\,.}
The field content of a single left or right semichiral multiplet is (3 scalars, 4 Weyl fermions, 1 chiral vector), where all fields are complex (see Appendix~\ref{Components of Semichiral fields} for definitions). As discussed below, we are interested in models containing pairs $(\Bbb X_{L},\Bbb X_{R})$ of left and right semichiral fields. Together, these  combine into the field content 
\eqsn{ 
(\Bbb X_{L},\Bbb X_{R}):\quad \, (\underbrace{\text{2 scalars, 4 Weyl fermions}}_{\text{physical}})\oplus \underbrace{\text{(4 scalars, 1 vector, 4 Weyl fermions)}}_{\text{auxiliary}}\,.
}

R-symmetry plays an important role in $\N=(2,2)$ theories. Classically, there is a $U(1)_V\times U(1)_A$ R-symmetry, acting on the superspace coordinates by
\eqsn{
U(1)_V:\quad \theta^{\pm} \rightarrow e^{i \alpha }\theta^{\pm}\,, \qquad \qquad U(1)_A: \quad \theta^{\pm} \rightarrow e^{\mp i \beta }\theta^{\pm}\,,
}
and on superfields as:
\eqsn{
U(1)_V:& \quad \Bbb X(x^{\pm},\theta^\pm, \bar \theta^\pm) \to e^{i q_V \alpha}\, \Bbb  X(x^{\pm}, e^{i \alpha }\theta^\pm,e^{-i \alpha } \bar \theta^\pm)\,,\\
U(1)_A:& \quad \Bbb X(x^{\pm},\theta^\pm, \bar \theta^\pm) \to e^{i q_A \beta}\, \Bbb  X(x^{\pm}, e^{\mp i \beta }\theta^\pm,e^{\pm i \beta } \bar \theta^\pm)\,,
}
where $q_{V,A}$ are the corresponding R-charges of the lowest component of the multiplet. The R-charge of the component fields is easily found from the definitions in Appendix~\ref{Components of Semichiral fields}.

\subsection{Linear Sigma Models}

We now discuss dynamics. Consider the simplest class of sigma models; linear sigma models. Already at this level there are some important differences with models for chiral or twisted chiral fields. In some respects semichiral models are more restricted, but in other respects they are less restricted. 
Consider first a single left semichiral field. A kinetic action would naturally be given by \cite{Buscher:1987uw}
\equ{
\int d^{4}\theta \, \Bbb{ \bar X}_L \Bbb{ X}_L\,.
}
However, it is easy to see that this action leads to linear kinetic terms rather than quadratic. The equations of motion for the components fields\footnote{We denote the $\N=0$ components of a left semichiral multiplet by $(X_L,F_L,M_{-+},\psi^l_{\pm},\bar \chi_{-}^{l},\bar \eta_{-}^{l},M_{--})$ and their conjugates, and for right semichiral multiplet by $(X_R,F_R,M_{+-},\psi^r_{\pm},\bar \chi_{+}^{r},\bar \eta_{+}^{r},M_{++})$ and their conjugates. See Appendix~\ref{Components of Semichiral fields} for definitions.} set $M_{-+}=\bar M_{-+}=\psi_+^l=\bar \psi_+^l=0$ and
\eqsn{
\partial_{++}X_L=\,& \partial_{++}\bar{X}_L=\partial_{++}M_{--}=\partial_{++}\bar M_{--}=0\,,\\
\partial_{++}\psi_-^l=\,&\partial_{++}\bar \psi_-^l= \partial_{++}\chi_-^l=\partial_{++}\bar \chi_-^l=0\,,
}
which describe two left-moving bosonic and two left-moving fermionic modes (of course, the analogous statement holds for a single right semichiral field). Although interesting, here we will focus on standard sigma models and  do not discuss these models further. 

To obtain a sigma model with standard kinetic terms one must consider an equal number of left and right semichiral fields, with couplings between them. For a single   left and right semichiral field the most general linear model one can write is
\equ{\label{matter action general}
\L_{matter}  = \int d^{4}\theta \, \left[ a \, \Bbb{ \bar X}_L \Bbb{ X}_L+b \, \Bbb{ \bar X}_R \Bbb{ X}_R + (c \,  \Bbb{ \bar X}_L \Bbb{ X}_R+c.c.) +( d \,  \Bbb{ X}_L \Bbb{ X}_R+ c.c.) \right]\,,
}
where $a,b$ are real and $c,d$ are complex parameters. As long as $c$ and $d$ are not both zero, one obtains standard kinetic terms  \cite{Buscher:1987uw}. Though $a,b\ne0$ can be scaled to $\pm1$, the parameters $c/\sqrt{ab}$ and $d/\sqrt{ab}$ cannot be absorbed by field redefinitions. The requirement of a positive definite kinetic term imposes minor conditions on the signs of $a,b$ and the range of the remaining parameters. Thus even for flat space, the action depends on nontrivial parameters; their significance is that they determine a choice of complex structures $J_\pm$ on the space (see \ref{J's flat space} for explicit expressions).

Semichiral models typically have fewer (rigid) flavor symmetries than models for chiral fields. For instance, for $c,d\neq 0$, (\ref{matter action general}) has no $U(1)$ isometries (although there are shift isometries). In the special case $c=0$ there is a $U(1)$ isometry acting by
\equ{\label{isometry c=0}
\Bbb{ X}_L \rightarrow e^{i q \lambda}  \Bbb{ X}_L\,,\qquad \Bbb{ X}_R \rightarrow e^{-i q \lambda}  \Bbb{ X}_R\,,}
where $\lambda$ is a real parameter. In the special case $d=0$ there is a $U(1)$ isometry acting by
\equ{\label{isometry d=0}
 \Bbb{ X}_L \rightarrow e^{i q \lambda}  \Bbb{ X}_L\,,\qquad \Bbb{ X}_R \rightarrow e^{i q \lambda}  \Bbb{ X}_R\,.}
More generally, if $(\Bbb X_L, \Bbb X_R)$ are valued in the Lie algebra of a group $G$, one can write a linear sigma model with an isometry if they are either in a representation $(\mathfrak R, \bar{ \mathfrak R})$ or $(\mathfrak R, \mathfrak R)$ of $G$.

Since we are interested in models with isometries, we will consider models in which either $c=0$ or $d=0$. In fact, these two cases are equivalent due to semichiral-semichiral duality \cite{Grisaru:1997ep}.

\subsection*{\it Semichiral-Semichiral duality}

In general, T-dualities change the target space geometry, 
but there are other kinds of dualities that amount to nothing more than a change of coordinates.
The oldest example is the duality between chiral and complex linear superfields \cite{Gates:1983nr}; another is a duality that relates semichiral superfields to themselves \cite{Grisaru:1997ep}. Both of these have the property that they
change the gauge charge $Q$ of the dualized multiplet to $-Q$. Starting with a model 
(\ref{matter action general}) with  $c=0$, one may dualize $\Bbb{X}_R$ by relaxing the semichiral condition and imposing it by a right-semichiral Lagrange multiplier term $\tilde{\Bbb X}_{R}\Bbb X_{R}+c.c.\,$ where $\tilde{\Bbb X}_{R}$ is a right semichiral Lagrange multiplier field. Integrating out $\tilde{\Bbb X}_{R}$ leads to the original model, while integrating out $\Bbb X_{R}$
we find a model with $\tilde d=0$ and $\tilde a=a -|d|^{2}/b,\tilde b=-1/b,\tilde c=\bar d/b$.

We stress that this duality does not change the geometry since it is simply a change of coordinates, and hence we call this a coordinate duality. Thus, without loss of generality one may consider models where all pairs of semichiral fields $(\Bbb X_{L}^i,\Bbb X_{R}^i)$ have charges either $(Q_i,-Q_i)$ or $(Q_i,Q_i)$. Depending on the type of computation, one or the other choice may be more convenient; in this paper we mostly make the former. 

\vspace{0.3cm}

Finally, we note that traditional superpotential terms for semichiral fields are not allowed; a term of the form $\int d^2 \theta f(\Bbb X_L,\Bbb X_R)$ breaks $\N=(2,2)$ supersymmetry. 

\subsection{Gauging and Vector Multiplets}

We now discuss the supersymmetric gauging of these models; this is where semichiral models are less restricted  than chiral models. For simplicity we describe the Abelian case (see Appendix~\ref{Vector Multiplets} for a review of the nonabelian case). 

To gauge an isometry while preserving supersymmetry, one promotes the rigid symmetry parameter $\lambda$ to a superfield which is compatible with the supersymmetric constraint on the fields the isometry acts on. To ensure the invariance of Lagrangians such as the ones just discussed under the local transformations, one introduces an appropriate vector multiplet.

\subsubsection{Vector and Twisted Vector Multiplets}

If the isometry acts on chiral and  semichiral fields, the symmetry parameter $\lambda$ is promoted to a chiral superfield $\Lambda$. The corresponding vector multiplet is the standard multiplet $V$ with gauge transformation $\delta_g V= i (\bar \Lambda-\Lambda)$.  The gauge-invariant field strength is given by 
\equ{\label{field strength usual vector}
\Sigma= i \, \bar{\Bbb D}_+ \Bbb D_- V
}
and is twisted chiral. The kinetic action for the vector is given by
\equ{\label{action usual vector}
\mathcal L_\Sigma=-\frac{1}{2e^{2}} \int d^{4}\theta \,  \bar {\Sigma} \Sigma\,.
}
A unitary gauge on the Higgs branch is found by setting a chiral matter field to its nonzero VEV.

If the isometry acts on twisted chiral and semichiral fields, $\lambda$ is promoted to a twisted chiral superfield $\tilde \Lambda$. The corresponding vector multiplet is the twisted vector $\tilde V$  with gauge transformation $\delta_g \tilde V= i (\bar{\tilde \Lambda}-\tilde{\Lambda})$. In this case the gauge-invariant field strength is given by
\equ{\label{field strength twisted vector}
\Theta= i\, \bar{\Bbb D}_+ \bar{\Bbb D}_- \tilde V\,,
}
which is chiral, and the kinetic  action is 
\equ{\label{action twisted vector}
\mathcal L_{\Theta} = \frac{1}{2e^{2}} \int d^{4}\theta \,  \bar {\Theta} \Theta\,.
}
In this case, a unitary gauge on the Higgs branch is found by setting a twisted chiral matter field to its nonzero VEV.

\subsubsection{Semichiral Vector Multiplet}

Finally, if the isometry acts only on semichiral fields one may promote $\lambda$ to a pair of left and right semichiral gauge parameters: 
\equ{
\Bbb X_{L}\rightarrow e^{i Q_{L} \Lambda_{L}} \Bbb X_{L}\,, \qquad \Bbb X_{R}\rightarrow e^{i Q_{R} \Lambda_{R}} \Bbb X_{R}\,.
}
The corresponding vector multiplet is the Semichiral Vector Multiplet (SVM) \cite{Lindstrom:2007vc,Lindstrom:2008hx}. The SVM is defined by $(V_{L},V_{R}, \Bbb V,\tilde{\Bbb V})$, with gauge transformations
\eqs{ \label{gauge transformations SVM}
\delta V_{L}=  i (\bar \Lambda_{L} - \Lambda_{L})\,, \quad  \delta V_{R}=  i (\bar \Lambda_{R} - \Lambda_{R})\,, \quad  i \delta \Bbb V = i ( \Lambda_{L} - \Lambda_{R})\,, \quad  i \delta \tilde{ \Bbb V} = i (\Lambda_{L} -\bar \Lambda_{R})\,;
}
here $V_{L,R}$ are real but $\Bbb V$ and $\tilde {\Bbb V}$ are complex. These vector multiplets are not completely independent, but satisfy the (gauge-invariant) relations
\equ{\label{relations SVM}
-\frac{1}{2}V'\equiv \text{Re} \, \Bbb{\tilde V}=\text{Re}\, \Bbb V \,,\qquad V_{R}= i\, (\Bbb{V}-\Bbb{\tilde V} )= i\,( \Bbb{\tilde{\bar V}} -  \bar{\Bbb{V}})\,,
\qquad V_{L}=i\,(\tilde{\bar{\Bbb{V}}}-
\Bbb{V})= i\,(\Bbb{\bar V} -  \tilde{\Bbb{V}})\,. 
}
There are two field strengths which are invariant under the full gauge symmetry (\ref{gauge transformations SVM}):
\equ{\label{definition field strengths}
  \tilde{\Bbb F}\equiv \bar{\Bbb D}_{+}\Bbb D_{-} \tilde{\Bbb V}\,, \qquad  \Bbb F\equiv\bar{\Bbb D}_{+}\bar{\Bbb D}_{-} \Bbb V\,;
} 
these are twisted chiral and  chiral, respectively. The kinetic action for the SVM is given by
\equ{ \label{action SVM}
\L_{SVM}=- \frac{1}{2e^{2}}\int d^{4}\theta
 \(\bar {\tilde{ \Bbb F}} \tilde{\Bbb F} -\bar {\Bbb F} \Bbb F \)\,.
}
The R-charges of $\Bbb F, \tilde{\Bbb F}$ follow from the definitions (\ref{definition field strengths}) and for reference are given in Table~\ref{R charges SVM}. 

\begin{table}[h]
\begin{center}
\begin{tabular}{|c|c|c|c|c|c|c|}
\hline
&~&~&~&~&~&~\\[-3ex]
 & $\Bbb D_{+}$&  $\Bbb D_{-}$& $\bar{\Bbb D}_{+}$& $\bar{\Bbb D}_{-}$ & $\Bbb F$ & $\tilde{\Bbb F}$\\
\hline
$q_V$ &$ -1$&$-1$ &$1$ & $1$ & $2$ & $0$ \\
\hline
$q_A$  & $1$& $-1$ &$-1$ &$1$ & $0$ & $-2$ \\
\hline
\end{tabular}
\end{center}
\caption{Axial and vector R-charges of fermionic derivatives and field strengths in the SVM.}
\label{R charges SVM}
\end{table}

Possible Fayet-Iliopoulos (FI) terms are 
\equ{ \label{FI terms SVM}
\L_{FI} =\(i \, t \int d^{2}\tilde \theta  \, \tilde{\Bbb F}+c.c.\)+ \(i \, s \int d^{2}\theta  \, \Bbb F+c.c.\) \,,
}
where $d^2\theta= d \theta^+ d\theta^-,\, d^2 \tilde \theta= d \bar \theta^- d \theta^+$ and $s,t$ are complex parameters. These terms are compatible with R-symmetry. From the definitions (\ref{definition field strengths}), they can also be written as the D-term\footnote{Here we have dropped the topological term $\sim \text{Im}(s-t) F_{01}$ which should be included.} $i\int d^4 \theta (t \tilde{\Bbb V}+s \Bbb V)+c.c.$.

When the isometry acts as in (\ref{isometry c=0}), the gauge-invariant matter Lagrangian is given by 
\equ{  \label{SVM gauged action opposite charge}
\L_{matter}=+\int d^{4}\theta
\left[ \Bbb{ \bar X}_L e^{QV_L } \Bbb{ X}_L + \Bbb{ \bar X}_R e^{-Q V_R} \Bbb{ X}_R+  \beta \, (\Bbb{ X}_L e^{- i Q \Bbb V} \Bbb{ X}_R+c.c. )  \right]\,,
}
where we have rescaled the fields in (\ref{matter action general}) to set $|a|=|b|=1$ and by a further field redefinition we take $d=\beta$ real and non-negative. Finally, the requirement of a positive definite kinetic term imposes $a=b=1$ and $\beta>1$.

Similarly, when the isometry acts as in (\ref{isometry d=0}) the gauge-invariant matter Lagrangian is given by 
\equ{ \label{SVM gauged action same charge}
\tilde \L_{matter}=-\int d^{4}\theta
\left[  \Bbb{ \bar X}_L e^{Q V_L } \Bbb{ X}_L + \Bbb{ \bar X}_R e^{Q V_R} \Bbb{ X}_R+ \alpha\,(\Bbb{ \bar X}_L e^{ iQ\bar{ \tilde{\Bbb V}}} \Bbb{ X}_R+c.c. )  \right]\,,
}
where $\alpha>1$ for a positive definite metric.
In both cases, a unitary gauge on the Higgs branch is found by setting a semichiral pair $\Bbb{ X}_L,\Bbb{ X}_R$ to its nonzero VEV.

\section{New GLSMs for Semichiral Fields}
\label{GLSMs for Semichiral Fields}

In this section we introduce a new vector multiplet that couples to semichiral fields and discuss various aspects of GLSMs including R-symmetry and twisted masses. 

\subsection{Constrained SVM}
\label{Constrained SVM}

We introduce a new \textit{constrained} semichiral vector multiplet or {\em CSVM}. As reviewed in Section~\ref{Background}, in the SVM there are two field strengths: $\Bbb F$ and $\Bbb{\tilde F}$, chiral and twisted chiral, respectively. The CSVM is obtained by constraining one of these field strengths (but not both) to vanish: 
 \equ{
\Bbb F=0\, \qquad \text{or} \qquad \Bbb{\tilde F}=0\,.
}
After a partial gauge-fixing, the first constraint reduces the SVM to the usual vector multiplet, while the second one reduces it to the twisted vector multiplet; however, the CSVM couples naturally to semichiral fields---its gauge parameters are semichiral and a unitary on the Higgs branch is still found by setting a semichiral pair $\Bbb{ X}_L,\Bbb{ X}_R$ to its nonzero VEV. As we discuss below, one may also constrain either of the fields strengths to be a non-zero constant, which is a way to introduce twisted masses. 

We first consider the case $\Bbb F=0$, which we  impose by a chiral Lagrange multiplier $\Phi$:
\equ{ \label{action CSVM}
\L_{CSVM}=- \frac{1}{2e^{2}}\int d^{4}\theta \(\bar {\tilde{ \Bbb F}} \tilde{\Bbb F} -\bar {\Bbb F} \Bbb F \)+\(i\int d^{2}\theta \,  \Phi\, \Bbb F+c.c.\)\,.
}
We wish to understand how the semichiral gauge parameters in (\ref{gauge transformations SVM}) are restricted to chiral parameters when we impose $\Bbb F =0$;
the constraint implies (locally) that $\Bbb V$ is pure gauge, $i.e.$,
\equ{ \label{chiral constraint}
i\Bbb V = i(\Lambda_{L}-\Lambda_{R})\,. 
}
{\em This is invariant} under a restricted gauge transformation $\delta_{c}$ with parameters
\equ{ 
\Lambda_{L}^{c}=\Lambda_{R}^{c}\equiv\Lambda \qquad \Rightarrow \qquad \text{$\Lambda$ chiral}\,.
}
Thus, there is a residual gauge symmetry associated to gauge transformations 
with a chiral parameter. From (\ref{gauge transformations SVM}) we have  
\equ{\label{gauge transformations constrained tilde V}
 -i \delta_{c}\,\tilde{ \Bbb V} = i (\bar\Lambda - \Lambda)\,,
 }
which is the gauge transformation for the usual vector multiplet. Now, because $\Bbb V$ is pure gauge, we can set $\Bbb V=0$ and using (\ref{relations SVM}) we have
\equ{ \label{solution VL VR}
 V_{L}=V_R= -i\,\tilde{\Bbb V}=i\,\bar{\tilde{\Bbb V}}\equiv V\,,
}
where we have introduced the real field $V$, which from (\ref{gauge transformations constrained tilde V}) transforms under the residual symmetry as  $\delta_{c}V = i(\bar \Lambda -\Lambda)$, so it is naturally identified with the usual vector multiplet. Furthermore, from the definition (\ref{definition field strengths}) and  (\ref{solution VL VR}), one sees that $\tilde{\Bbb F}=  i \, \bar{\Bbb D}_{+}\Bbb D_{-} V$ coincides with the usual field strength (\ref{field strength usual vector}) and the action (\ref{action SVM}) reduces to the standard  action  (\ref{action usual vector}). Thus, we have shown that imposing the constraint $\Bbb F=0$ on the SVM effectively reduces it to the usual vector multiplet after partial gauge-fixing. At this point, one could simply talk about the usual vector multiplet rather than the CSVM. However, the CSVM description is particularly useful when coupling to matter fields, as we discuss below.

One may instead constrain the twisted chiral field strength $\Bbb{\tilde F}$ to vanish by introducing a twisted chiral Lagrange multiplier $\chi$:
\equ{  \label{action twisted chiral constraint}
i\int d^{2}\tilde \theta \,  \chi\, \Bbb{\tilde F}+c.c.\,. 
}
The equation of motion for $\chi$ imposes $\Bbb{\tilde F}=0$, or
\equ{ \label{twisted chiral constraint}
i \tilde{ \Bbb V} = i (\Lambda_{L} -\bar \Lambda_{R})\,,
}
which is invariant under the residual gauge transformation $\delta_{tc}$ with parameters
\equ{ 
\Lambda_{L}^{tc}=\bar{\Lambda}_{R}^{tc}\equiv \tilde \Lambda \qquad \Rightarrow  \qquad \text{$\tilde \Lambda$ twisted chiral}\,.
}
This remaining gauge invariance is gauged by the twisted vector multiplet $\tilde V=-i \Bbb V$ with gauge transformation $\delta_{tc}\tilde V=i (\bar{\tilde \Lambda}-\tilde \Lambda)$. With this constraint $\Bbb  F$ becomes the field strength (\ref{field strength twisted vector}) and the action (\ref{action SVM}) reduces to (\ref{action twisted vector}).

Thus, we have shown that the Abelian SVM can be constrained to effectively become either the usual vector or the twisted vector multiplet.\footnote{In addition to the SVM, there is another novel vector multiplet, the LVM (Large Vector Multiplet), which couples to chiral and 
twisted chiral fields simultaneously \cite{Lindstrom:2007vc,Lindstrom:2008hx}. One might
wonder if it can be constrained as well; it seems unlikely, however, as the gauge parameters of the LVM
are chiral and twisted chiral, and one cannot make a consistent off-shell constraint relating them.} As we discuss below, this perspective is useful to study the coupling of semichiral fields to the various vector multiplets, introduce twisted masses, and describe the target space geometry compactly by a generalized potential in $\N=(2,2)$ superspace.

\subsection{Coupling to Matter}

Here we discuss the coupling of the CSVM to semichiral matter multiplets. Consider a pair of semichiral fields with opposite charges, coupled to the SVM constrained by $\Bbb F=0$:
\eqs{ \nonumber
\L=\,&- \frac{1}{2e^{2}}\int d^{4}\theta \(\bar {\tilde{ \Bbb F}} \tilde{\Bbb F} -\bar {\Bbb F} \Bbb F \)  +\(i \int d^{2}\theta \,  \Phi\, \Bbb{ F}+c.c.\)+\(i \, t \int d^{2}\tilde \theta  \, \tilde{\Bbb F}+c.c.\)\\ \label{SVM gauged action opp charge constraint}
&\,+\int d^{4}\theta
\left[ \Bbb{ \bar X}_L e^{Q V_L } \Bbb{ X}_L + \Bbb{ \bar X}_R e^{ -QV_R} \Bbb{ X}_R+  \beta \,( \Bbb{ X}_L e^{- i Q \Bbb V} \Bbb{ X}_R+c.c. )  \right]  \,,
}
where $t$ is an FI parameter (the parameter $s$ in (\ref{FI terms SVM}) can be set to zero by a shift in $\Phi$). As discussed, the equation of motion for $\Phi$ sets $\Bbb F=0$. Choosing the gauge $\Bbb V=0$ and using  (\ref{solution VL VR}), the action becomes
\equ{  \label{semis opposite charges gauged vector}
\L=\int d^{4}\theta
\left[ - \frac{1}{2e^{2}}\bar {\Sigma} \Sigma+   \Bbb{ \bar{ X}}_L e^{QV } \Bbb{ X}_L + \Bbb{ \bar{ X}}_R e^{-QV} \Bbb{X}_R+  \beta \, ( \Bbb{  X}_L \Bbb{ X}_R+c.c. )  \right]+\(i \, t \int d^{2}\tilde \theta  \, \Sigma +c.c.\)\,,
}
which is invariant under the residual, chiral, gauge symmetry
\equ{ 
\delta  V= i(\bar{\Lambda}- \Lambda)\,,\qquad \Bbb X_{L}\rightarrow e^{i Q \Lambda} \Bbb X_{L}\,, \qquad \Bbb X_{R}\rightarrow e^{-i Q \Lambda} \Bbb X_{R}\,.
}
Thus, the model (\ref{SVM gauged action opp charge constraint}) effectively reduces to a gauge theory of semichiral fields coupled to the usual vector multiplet. 

Coupling semichiral fields to the SVM constrained by $\Bbb{\tilde F}=0$ is similar: 
one finds 
\equ{ 
\L=\int d^{4}\theta
\left[  \frac{1}{2e^{2}} \, \bar {\Theta} \Theta+   \Bbb{ \bar{X}}_L e^{Q \tilde V } \Bbb{ X}_L + \Bbb{ \bar{ X}}_R e^{Q\tilde V} \Bbb{ X}_R+   \beta \,(\Bbb{ {X}}_L e^{Q \tilde V} \Bbb{X}_R+c.c. )  \right]+\(i \, s \int d^{2}\theta \, \Theta+c.c.\)\,,
}
where $s$ is an FI parameter. This is invariant under the residual, twisted chiral, gauge symmetry
\equ{
\delta \tilde V= i(\bar{\tilde \Lambda}-\tilde \Lambda)\,,\qquad \Bbb X_{L}\rightarrow e^{i Q \tilde \Lambda} \Bbb X_{L}\,, \qquad \Bbb X_{R}\rightarrow e^{-i Q \bar{\tilde \Lambda}} \Bbb X_{R}\,.
}
Thus, the gauge theory with matter effectively reduces to a gauge theory of semichiral fields coupled to the twisted vector multiplet. The same analysis can be repeated for the case of opposite charges. 

Since by constraining $\Bbb F=0$ we arrive at a usual vector multiplet, one might wonder if the CSVM can couple to ordinary chiral matter; however, since the original coupling involves semichiral gauge parameters, this is not possible. Of course, when the model includes charged chiral fields, one can directly introduce a standard vector multiplet and add to (\ref{semis opposite charges gauged vector}) the kinetic terms for the chiral fields:
\equ{\label{kinetic term chirals}
\int d^4 \theta \, \bar \Phi^a e^{q_a V} \Phi^a \,.
}
The corresponding statement holds for the CSVM with  $\Bbb{\tilde F}=0$ and the addition of charged twisted chiral fields.

\subsection*{\it Multiflavor case}

The generalization to the multiflavor case is straightforward. Consider $N_{s}$ pairs of semichiral fields  $( \Bbb{ X}_L^{i} , \Bbb{ X}_R^{i} )$, $i=1,...,N_{s}$, charged under a single single $U(1)$.  If we take the left and right semichiral field in each pair to have opposite charges the action is 
\eqs{ \nonumber
\L=\,&- \frac{1}{2e^{2}}\int d^{4}\theta \(\bar {\tilde{ \Bbb F}} \tilde{\Bbb F} -\bar {\Bbb F} \Bbb F \)  +\(i \int d^{2}\theta \,  \Phi\, \Bbb{ F}+c.c.\)\\ \label{SVM gauged action same charge Nf}
&\,+\int d^{4}\theta
\left[ \Bbb{ \bar X}_L^{i} e^{Q_{i}V_L } \Bbb{ X}_L^{i} + \Bbb{ \bar X}_R^{i} e^{-Q_{i}V_R} \Bbb{ X}_R^{i}+   \beta_i \,(\Bbb{ X}_L^{i} e^{ -iQ_{i}\Bbb V } \Bbb{ X}_R^{i}+c.c. )  \right]  \,,
}
where $\beta_i>1$. Integrating out $\Phi$ and using (\ref{solution VL VR}) we have
\equ{  \label{many semis chiral gauging}
\L=\int d^{4}\theta
\left[ - \frac{1}{2e^{2}}\bar {\Sigma} \Sigma+   \Bbb{ \bar{ X}}_L^{i} e^{Q_{i}V } \Bbb{ X}_L^{i} + \Bbb{ \bar{ X}}_R^{i} e^{-Q_{i}V} \Bbb{  X}_R^{i}+   \beta_i \,(\Bbb{{ X}}_L^{i} \Bbb{  X}_R^{i}+c.c. )  \right]\,.
}

\subsubsection*{\it Diagonalizing the semichiral GLSM Lagrangian}

More generally, for compatible charges, we could consider various kinds of flavor mixing terms. However, it is straightforward to show that we can eliminate them by a combination of field redefinitions and coordinate dualities of the type described at the end of Section 2; the argument is as follows:
First, note that the only flavors that can interact (other than through gauge-interactions) are fields with the same or opposite charges. So without loss of generality, we can restrict our attention to a block with charges $\pm Q$. We distinguish the hermitian terms of type $a,b$ in (\ref{matter action general}) and the complex terms of type $c,d$, where now $a,b$ are nondegenerate Hermitian matrices obeying certain positivity conditions and $c,d$ are complex matrices obeying a certain nondegeneracy condition. The $a,b,c$ matrices are further block-diagonal with separate blocks that do not mix fields with charges $+Q$ and $-Q$, whereas, the $d$ matrix is block off-diagonal, with entries only for fields with opposite charges. We use the coordinate dualities to ensure that all the left semichiral fields $\Bbb{X}_L$ have charge $+Q$ and all the right semichiral fields $\Bbb{X}_R$ have charge $-Q$; then $c=0$ and $d$ is nondegenerate.
At this stage, it is easy to state the positivity condition on $a,b$: they have only positive eigenvalues. Next, we temporarily absorb $d$ into a redefinition of $\Bbb{X}_L$;
Now we can transform $\Bbb{X}_L\to \Bbb{X}_L M$ and 
$\Bbb{X}_R\to M^{-1} \Bbb{X}_R$ for an arbitrary invertible matrix $M$; this allows us to transform $a$ into the identity matrix (recall that we have assumed its eigenvalues are positive). So for the moment we have $a=d=\Bbb{I}$. This is preserved by arbitrary unitary transformations $\Bbb{X}_L\to \Bbb{X}_L U$ and $\Bbb{X}_R\to U^{-1} \Bbb{X}_R$, which allow us to diagonalize the Hermitian matrix $b$; finally we rescale the right semichiral fields $\Bbb{X}_R$ to set $b=\Bbb{I}$; now $d$ is a positive diagonal matrix; definiteness of the metric requires that all its eigenvalues $\beta_i>1$.

For generic finite values of the parameters $\beta_i$ the rigid flavor symmetry of the system is $U(1)^{N_s-1}$ (we will discuss R-symmetry shortly). If some $\beta_i$'s are equal the symmetry is enhanced. We will discuss symmetries and special cases in more detail later.

\subsection*{\it Twisted masses}

In the presence of flavor symmetries, it is usually possible to introduce twisted masses by gauging the flavor symmetry and setting the associated field strength to a constant background. As discussed, semichiral models have less flavor symmetries than their chiral counterparts and some twisted masses are not possible. Nonetheless, due to the enlarged gauge symmetry of the SVM one may introduce a new kind of mass parameter, which does not require additional flavor symmetries. This is achieved by constraining one of the field strengths in the SVM (which may be the one gauging the semichiral isometry) not to vanish but instead to be a constant $M$. This can be imposed by the term
\equ{ \label{twisted mass chiral}
 i\int d^{2}\theta \, \Phi\, (\Bbb F- M) +c.c.\,.
}
Integrating out $\Phi$ sets $\Bbb F=M$, or $\Bbb V =  {\Bbb V}_{M}$ with $ \bar{\Bbb D}_{+} \bar{\Bbb D}_{-}{\Bbb V}_{M}=M$, and we can write 
\eqs{ \label{solution VL VR twisted mass}
V_{L}=\,V+ p \, \text{Im} {\Bbb V}_{M}\,,\qquad V_{R}=\,V-(2-p)\, \text{Im} {\Bbb V}_{M}\,,\qquad V'=-2\ \text{Re}\Bbb{V}_M\,,
}
where $p$ is a real parameter, $V$ is the usual vector multiplet, and $\tilde{\Bbb F}= \bar{\Bbb D}_{+}\Bbb D_{-} \tilde{\Bbb V}= i \, \bar{\Bbb D}_{+}\Bbb D_{-} V $. This should be understood as a gauged linear sigma model with a twisted mass for semichiral fields coupled to a CSVM.

Clearly, one could instead constrain the twisted chiral field strength to a constant $\tilde{\Bbb F}=\tilde M$ by the term
\equ{ \label{twisted mass twisted chiral}
i\int d^2 \tilde \theta \, \chi \, (\tilde{\Bbb F}-\tilde M)+c.c.\,.
}

Due to the R-charges of the SVM field strengths (see Table~\ref{R charges SVM}), giving $\Bbb F$ a non-zero VEV breaks the vector R-symmetry, but not the axial one, while giving  $\tilde{\Bbb F}$ a non-zero VEV breaks the axial R-symmetry, but not the vector one.

If one doesn't care about R-symmetry, one can introduce both kinds of twisted masses simultaneously using both the $\Phi$ and $\chi$ couplings in (\ref{twisted mass chiral}.\ref{twisted mass twisted chiral}); this does not give rise to a gauge coupling, as the semichiral vector multiplet is completely constrained, and can be interpreted as a T-duality transformation that replaces the semichiral pair $\Bbb X_L,\Bbb X_R$ with the chiral and twisted chiral fields $\Phi,\chi$; after this T-duality, the twisted mass terms become ordinary and twisted linear superpotential terms.

\subsection{Relation to complex linear superfields}
\label{Relation to complex linear superfields}
We have shown that gauge fixing the semichiral gauge invariance for a CSVM gives the usual vector multiplet coupled to semichiral fields. Naively, this is very strange--chiral gauge parameters cannot gauge away semichiral fields, {\it i.e.}, one cannot use them go to a unitary gauge. To understand this, instead of the partial gauge fixing of the CSVM to the usual vector multiplet, we can keep all the fields of the CSVM; then we can choose a unitary gauge, {\it e.g.},  $\Bbb X_L=\Bbb X_R=1$, and impose it in (\ref{SVM gauged action opp charge constraint}):
\eqs{ \nonumber
\L=\,&- \frac{1}{2e^{2}}\int d^{4}\theta \(\bar {\tilde{ \Bbb F}} \tilde{\Bbb F} -\bar {\Bbb F} \Bbb F \)  +\(i \int d^{2}\theta \,  \Phi\, \Bbb{ F}+c.c.\)\\ \label{Unitary gauge action}
&\,+\int d^{4}\theta
\left[  e^{Q V_L } +  e^{ -QV_R}+ \beta \, ( e^{ -i Q\Bbb V} +c.c. )  \right]  \,.
}
We would like to derive this directly from the effective vector multiplet coupling (\ref{semis opposite charges gauged vector}).
This is accomplished by shifting $V$ in (\ref{semis opposite charges gauged vector}) such that
\equ{\label{rescale V}
 \Bbb{ \bar X}_L e^{Q V } \Bbb{ X}_L ~ \to~  e^{Q V }\,.}
Then (\ref{semis opposite charges gauged vector}) can be rewritten as:
\equ{\label{action with complex linear}  
\L=\int d^{4}\theta
\left[  -\frac{1}{2e^{2}}(\bar {\Sigma}+ \Bd_+ \bar \Bd_- \Xi)( \Sigma+ \bar \Bd_+ \Bd_- \bar \Xi)+    e^{QV } + e^{-Q(V+i \Xi- i \bar\Xi)}+  \beta \, (e^{- i Q\Xi}+c.c. )  \right]\,,
}
where we defined
\equ{\label{def Xi} 
e^{- i Q\Xi} = \Bbb X_R \Bbb X_L
}
and $\Xi$ is a complex linear superfield because
$\bar\Bd_+\bar\Bd_-(\ln{{\Bbb X}_R}+\ln{{\Bbb X}_L})=0$.
We now dualize $\Xi$ to a chiral multiplet $\Phi$ \cite{Gates:1983nr} by relaxing the constraint on $\Xi$ and adding a chiral Lagrange multiplier $\Phi$:
\eqs{\nonumber 
\L=\int d^{4}\theta
\Big[  -\frac{1}{2e^{2}}(\bar {\Sigma}+ \Bd_+ \bar \Bd_- \Xi)( \Sigma+ \bar \Bd_+ \Bd_- \bar \Xi)+    e^{QV } + e^{-Q(V+i \Xi- i \bar\Xi)}+ \beta \, (  e^{- i Q\Xi}+c.c. )\\ +(i \Phi \, \Xi + c.c.) \Big]\,.
\label{complex linear dual charges gauged vector}
}
Finally, identifying $V_L=V$, $V_R=V+i \Xi- i \bar\Xi$, $\Bbb V=
\Xi$, and using (\ref{relations SVM}), (\ref{complex linear dual charges gauged vector}) becomes
\equ{\label{unitary gauge action half SVM}
\L=\int d^{4}\theta
\left[ - \frac{1}{2e^{2}}\bar{\tilde{\Bbb F}}\tilde{\Bbb F}+e^{QV_L } + e^{-Q V_R}+  \beta \, (e^{- i Q \Bbb V}+c.c. ) +( i \Phi \, \Bbb V + c.c.) \right]\,,
}
which is almost (\ref{Unitary gauge action}), except that the term $\bar{\Bbb F}{\Bbb F}$ in the SVM action does not appear in (\ref{unitary gauge action half SVM}); if we shift $\Phi\to\Phi-
\frac{i}{4e^2}\bar\Bd_+\bar\Bd_- \Bd_+\Bd_-\bar\Xi$ we match (\ref{Unitary gauge action}) exactly.
 
\subsection{R-Symmetry}
\label{R-Symmetry}

The classical R-symmetry may become anomalous at the quantum level. In semichiral models, however, we find that the anomaly is automatically cancelled within each pair of semichiral fields. This may not be too surprising, given that the GLSMs at hand are continuous deformations of anomaly-free theories for chiral fields, as we show in Section~\ref{Low energy behavior}. Nevertheless, it is instructive to see how anomaly cancellation occurs in semichiral models.

\begin{figure}[h]
\centering
\includegraphics[]{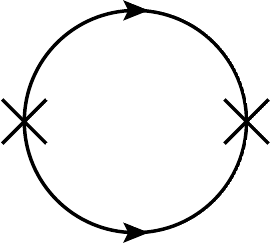}
\caption{One loop diagram contributing to the R-symmetry anomaly. One insertion represents the gauge current, the second the R-current, and fermions are running in the loop.}
\label{Anomaly}
\end{figure}

Possible R-symmetry anomalies arise from the one-loop diagram in Figure~\ref{Anomaly} and are proportional to
\eqs{\label{anomalies}
{\mathcal A}_V\propto \sum_{\text{fermions}} \gamma^3 Q_G \, q_V\,,\qquad \qquad {\mathcal A}_A\propto \sum_{\text{fermions}} \gamma^3 Q_G \, q_A\,,
}
where $\gamma_{3}$ is the chirality matrix, $Q_{G}$ is the gauge charge and $q_{V},q_{A}$ are the vector and axial R-charge, respectively. 

Let us first review the usual case of chiral and twisted chiral multiplets. In a chiral multiplet, there are two Weyl fermions $(\psi_+,\psi_-)$ with chiralities $\gamma_{3}:(1,-1)$, vector R-charges $q_{V}:(1,1)$, and axial R-charges $q_{A}:(1,-1)$.  Thus, from (\ref{anomalies}) it follows that the vector anomaly cancels automatically, but there is an axial anomaly proportional to the gauge charge $Q_{G}$. On the other hand, in a twisted chiral multiplet there are two Weyl fermions $(\tilde \psi_+,\tilde \psi_-)$ with chiralities $(1,-1)$, vector R-charges $(1,-1)$, and axial R-charges $(1,1)$.  Thus, the axial anomaly cancels automatically, but there is a vector anomaly proportional to $Q_{G}$. When there are several multiplets with gauge charges $Q_a$, the condition for anomaly cancellation in either multiplet is $\sum_a Q_a=0$.

Semichiral fields contain various fermions, which we denote by $(\psi^l_\pm,\chi_-, \eta_-)$ and  $(\psi^r_\pm,\chi_+, \eta_+)$ for left and right semichiral fields, respectively (see Appendix~\ref{Components of Semichiral fields} for definitions). However, they are not all propagating and therefore they do not all contribute to the one-loop diagram in Figure \ref{Anomaly}. To see which fermions contribute, consider for concreteness the case of opposite charges and the CSVM with $\Bbb F=0$, although the final conclusion is independent of these choices.  Reducing the action (\ref{SVM gauged action opp charge constraint}) to components, we see that the fermionic fields $\eta_{\pm}, \bar \eta_\pm$ appear only as
\equ{
\L=...+\eta^+\psi_+^l+\eta^-\psi_-^r + \bar \psi_l^-\bar \eta_-+\psi_r^+\bar \eta_+-\beta(\eta^+\chi_+ +\eta^-\chi_- + \bar \eta^+\bar \chi_++\bar \chi^+\bar \eta_+)\,,
}
which are simply Lagrange multipliers imposing $\chi_+=\tfrac{1}{\beta}\psi_+^l$ and $\chi_-=\tfrac{1}{\beta}\psi_-^r$, and similarly for the barred components. Thus, the only fermionic fields remaining are $(\psi^l_+,\psi^l_-)$ and $(\psi^r_+,\psi^r_-)$, which after simple field rescalings, have a canonical kinetic term and coupling to the vector field. Each of these fermions have chiralities $\gamma_{3}:(1,-1)$, vector R-charges $q_{V}:(1,1)$, and axial R-charges $q_{A}:(1,-1)$, but opposite gauge charges. Thus,
\equ{
{\mathcal A}_V={\mathcal A}_A=0\,.
}

Anomaly cancellation can also be seen in the case of semichiral fields of equal charges;  the Lagrange multipliers $\eta_\pm$ impose a different constraint among the remaining fermions, but the final conclusion is the same (as it should since these theories are related by a change of coordinates). The same analysis holds for the CSVM with $\tilde{ \Bbb F}=0$. Thus, the R-symmetry anomaly always cancels in the semichiral sector. If an addition to semichiral fields there are charged chiral or twisted chiral fields, coupled to the corresponding vector multiplets, they will give their usual contribution to the R-symmetry anomaly. 

The anomalies can also be computed using the Atiyah-Singer index theorem. The two complex structures $J_{\pm}$ induce two different decompositions of the complexified tangent bundle \cite{Kapustin:2004gv} $T \mathcal{M}_{\Bbb C}\simeq T_+^{1,0}\oplus T_+^{0,1}\simeq T_-^{1,0}\oplus T_-^{0,1}$ and the conditions for anomaly cancellation for a generalized K\"{a}hler manifold are
\eqsn{
U(1)_{V}:\quad c_{1}(T_{-}^{1,0})-c_{1}(T_{+}^{1,0})=&\,0\\
U(1)_{A}:\quad c_{1}(T_{-}^{1,0})+c_{1}(T_{+}^{1,0})=&\,0\,.
}
As discussed in \cite{Kapustin:2004gv}, these conditions are weaker than the Generalized Calabi-Yau condition described by Hitchin \cite{Hitchin:2004ut} and Gualtieri \cite{Gualtieri:2003dx}. 

\section{Low energy behavior}
\label{Low energy behavior}

In this section we describe the classical space of vacua of various GLSMs for semichiral fields.

As usual, the space of classical vacua is determined by the space of zero's of the classical potential $U$, modulo gauge transformations. The structure of the moduli space in GLSMs involving the SVM or the CSVM is very different so we analyze them separately. For simplicity, we restrict ourselves to a $U(1)$ gauge group.\footnote{The generalization to nonabelian gauge groups in the case of the SVM is straightforward. The case of the nonabelian CSVM will be studied elsewhere.}

Instead of reducing to $\N=0$ components directly to compute the classical scalar potential, it is convenient to partially reduce to $\N=(1,1)$ superspace first; this allows an easy comparison of the GLSMs at hand to standard GLSMs for chiral fields. We denote the $\N=(1,1)$ gauge covariant fermionic derivatives by $\D_{\pm}$ (see Appendix~\ref{Reduction to $(1,1)$}). The basic multiplets in $\N=(1,1)$ SUSY are described by unconstrained bosonic and fermionic superfields and the vector multiplet with superfield strength $f$. Chiral and twisted chiral superfields both reduce to a single complex bosonic superfield and the vector multiplet reduces to a $\N=(1,1)$ vector multiplet $f$ and a real bosonic superfield $\sigma$. Semichiral multiplets $\Bbb X_L, \Bbb X_R$ reduce to complex bosonic superfields $X_L,X_R$, and auxiliary fermionic superfields. The SVM reduces to $(f,\sigma_1,\sigma_2,\sigma_3)$, where $\sigma_I$ are real bosonic superfields. 

\subsection{SVM}

Consider the action
\equ{
\L= \L_{SVM}+\L_{FI}+ \L_{matter}\,,
}
where each term is given in (\ref{action SVM}), (\ref{FI terms SVM}) and (\ref{SVM gauged action opposite charge}). Reducing the Lagrangian to $\N=(1,1)$ superspace gives (see Appendix~\ref{Reduction to $(1,1)$} for details)
\eqs{
\L= \L_{SVM}+\int \D_+ \D_- \,\( \frac{1}{2}\(g_{\mu \nu}+b_{\mu \nu}\)\D_+X^{\mu}\D_- X^{\nu}+2 i\, \sigma_I( \mu_I-r_I)\)\,,
\label{action 1,1 GLSM SVM general}
}
where $\L_{SVM}$ is given in (\ref{kinetic action SVM 1,1}), $X^\mu=(X_L, \bar X_L, X_R,\bar X_R)$ and $g,b$ are the flat-space metric and $b$-field of the ungauged case (set $a=b=1, d=\beta$ in \ref{g and b flat space}):
\equ{g=4
\left(
\begin{array}{cccc}
 0 &  1 & \frac{1}{\beta} & 0 \\
  1 & 0 & 0 & \frac{ 1}{\beta} \\
 \frac{ 1}{\beta} & 0 & 0 &  1 \\
 0 & \frac{ 1}{\beta} &  1 & 0 \\
\end{array}
\right)\,,\qquad  b= 2\(\tfrac{2}{\beta}-\beta\) \left(
\begin{array}{cccc}
 0 & 0 & 1 & 0 \\
 0 & 0 & 0 & 1 \\
  -1 & 0 & 0 & 0 \\
 0 &  -1 & 0 & 0 \\
\end{array}
\right)\,,
\label{g and b flat space beta}
}
and we defined the functions
\eqs{\nonumber
\mu_{1}\equiv \,& \bar X_{L}X_{L}+ \bar X_{R}X_{R}+\beta  (X_{L} X_{R}+\bar X_{L}\bar X_{R})\,,\\ \label{definition moment maps}
\mu_{2}\equiv \,&-( \bar X_{L}X_{L}- \bar X_{R}X_{R})\,,\\ \nonumber
\mu_{3}\equiv \,&- \frac{i \beta}{2}\(X_{L}X_{R}-\bar X_{L}\bar X_{R}\)\,,
}
and $r_{I}$ are real FI parameters defined as $r_1=2 \,\text{Re}\,s, r_2=-2\, \text{Re}\,t, r_3=\text{Im}(s+t)$. 

The Lagrangian (\ref{action 1,1 GLSM SVM general}) is not the standard way of writing the matter $\N=(1,1)$ Lagrangian; in 
\cite{Hull:1989jk,Jack:1989ne,Kapustin:2006ic} it is written in a way that makes clear the invariance of the action under $b$-field transformations---see discussion below (\ref{definition moment maps app}).

Further reducing to $\N=0$ and integrating out the auxiliary fields leads to the scalar potential (see Appendix~\ref{Reduction to N=0 and scalar potential}): 
\equ{\label{scalar potential SVM}
 U=2 e^{2} (\mu_{1}-r_{1})^{2}+2 e^{2}(\mu_{2}-r_{2})^{2}+4 e^{2}(\mu_{3}-r_{3})^{2}+\beta^{2}\(|\sigma|^{2}+\tfrac{1}{\beta^{2}-1}|\tilde \sigma|^{2}\) \frac{1}{2}|X|^2\,,
}
where $|X|^2\equiv X^{\mu}g_{\mu \nu}X^{\nu}$, with $g$ given in (\ref{g and b flat space beta}) and $\sigma=\Bbb F|$ and $\tilde \sigma =\tilde{\Bbb F}|$ are the complex scalars in the SVM.

There are two branches: the Coulomb branch and the Higgs branch.

\begin{enumerate}[(a)]
\item The Coulomb branch is parametrized by the VEVs of $\sigma$ and $\tilde \sigma$ and $X_{L}=X_{R}=0$. This branch exists only for $r_{1}=r_{2}=r_{3}=0$.  

\item The Higgs branch is given by $\sigma=\tilde \sigma=0$, and the space of solutions to
\equ{\mu_{I}=r_{I}\,,  \qquad \qquad I=1,2,3\,\,,}
modulo $U(1)$ gauge transformations. 
\end{enumerate}

In this simple model with a single pair of semichiral fields the Higgs branch is a point. The generalization to the multiflavor case with fields $(\Bbb X_L^i, \Bbb X_R^i)$, $i=1,...,N_s$ (and corresponding  parameters $\beta_i$) is straightforward, in which case the complex dimension of the Higgs branch is $2N_s-2$. For generic values of $\beta_i$ the rigid symmetry of the system is $U(1)^{N_s-1}\times U(1)_A \times U(1)_V$ and the geometry is generalized K\"{a}hler. If some $\beta_i$ coincide the rigid symmetry is enhanced. In the special case in which all $\beta_i$ coincide, the system has an additional rigid $SU(N_s)$ symmetry and the geometry becomes hyperk\"{a}hler \cite{Crichigno:2011aa} (for $N_{s}=2$, it is the Eguchi-Hanson solution). Thus, these models provide continuous deformations of hyperk\"{a}hler geometries; we leave a detailed study of these spaces for future work.

It may also be interesting to study the geometry on the SVM Coulomb branch, as in the case of the $\N=(4,4)$ vector multiplet \cite{Diaconescu:1997gu}. Here we focus on the geometry of the Higgs branch.

\subsection{Constrained SVM}

Now we consider a GLSM for a pair of semichiral fields with opposite charges, coupled to the CSVM with a twisted mass and FI parameter: 
\equ{
\L=\L_{SVM}+\L_{\Phi;M}+\L_{FI}+\L_{matter}\,,
}
where each term is given in (\ref{action SVM}), (\ref{FI terms SVM}),  (\ref{SVM gauged action opposite charge}), and $\L_{\Phi;M}$ is given by (\ref{twisted mass chiral}). By a shift in $\Phi$ we set $s=0$ in $\L_{FI}$ and the only FI parameter left is $t\equiv \tfrac12(\xi-i \tfrac{\theta}{2 \pi})$. As discussed below (\ref{Lphi 1,1}), in $\N=(1,1)$ language, constraining the SVM corresponds to setting $\sigma_1=2\, \text{Re}\, M$ and $\sigma_3=f-4 \,\text{Im}\,M$, while constraining  $\tilde{\Bbb F}=\tilde M$ corresponds to setting $\sigma_2=2\, \text{Re}\, \tilde M$ and $\sigma_3=-f+4\, \text{Im} \,\tilde M$. Using this in (\ref{action 1,1 GLSM SVM general}) gives
\eqs{\nonumber
\L=\L_{gauge}+\int \D_+ \D_- \,&  \Big( \frac{1}{2}(g_{\mu \nu}+b_{\mu \nu})\D_+X^{\mu}\D_- X^{\nu}+2 i\, \sigma_2( \mu_2+\xi)\\
\,&+2 i\, (f \mu_3+2 \mu_1 \text{Re}\,M -  4 \mu_3 \, \text{Im}\, M )\Big)\,,
\label{reduction 1,1 GLSM CSVM} 
}
where $\L_{gauge}$ is the action for the standard vector.

Finally, reducing  to $\N=0$ components, and integrating out the auxiliary fields we obtain (see Appendix~\ref{Reduction to N=0 and scalar potential})
\equ{
U=2 e^{2}(\mu_{2}+\xi)^{2}+\beta^{2}\(|M|^{2}+\tfrac{1}{\beta^{2}-1}|\tilde \sigma|^{2}\) \frac{1}{2}|X|^2\,.
}

The space of supersymmetric vacua depends on the values of the parameters $\xi$ and $M$.   

\begin{enumerate}[(a)]

\item For  $M\neq 0$ and $\xi\neq0$ supersymmetry is broken and there is no moduli space.
\item For $ M\neq 0$ and $\xi=0$ there is only the Coulomb branch paramaterized by $\tilde \sigma$, and $X_{L}=X_{R}=0$. 

\item For $M=0$ there can be  two branches. The Higgs branch is given by $\tilde \sigma=0$ and the space of solutions to 
\equ{
   |X_{L}|^2- |X_{R}|^2=\xi\,,
}
modulo gauge transformations. For $\xi=0$ there is also a  Coulomb branch, parametrized by the VEV of $\tilde \sigma$ and $X_{L}=X_{R}=0$. \label{Higgs branch bullet point} 
\end{enumerate}

The case of the CSVM with $\Bbb{\tilde F}=\tilde M$ is completely analogous (see Appendix~\ref{Reduction to N=0 and scalar potential}). 
\medskip

From now on we will focus on the Higgs branch in (c). Setting $M=0$ in (\ref{reduction 1,1 GLSM CSVM}) we can rewrite the Lagrangian (up to total derivatives) as 
\equ{\label{reduction 1,1 GLSM CSVM M=0}
\L= \L_{gauge}+\int \D_+ \D_-\( \frac{1}{2}(g_{\mu \nu}+b_{\mu \nu}')\D_+X^{\mu}\D_- X^{\nu}+2 i \sigma_2( \mu_2+\xi)\)\,,
}
where we defined
\equ{\label{b prime}
b'= \tfrac{4}{\beta} \left(
\begin{array}{cccc}
 0 & 0 & 1 & 0 \\
 0 & 0 & 0 & 1 \\
  -1 & 0 & 0 & 0 \\
 0 &  -1 & 0 & 0 \\
\end{array}
\right)\,.
}
We note that in the special limit $\beta \to \infty$ the metric $g$ given in  (\ref{g and b flat space beta}) becomes the canonical one, the gauged WZ term $b'$ in (\ref{b prime}) vanishes, and the Lagrangian coincides with that of chiral fields with charges $(1,-1)$ under $U(1)$, gauged by the usual vector multiplet (cf. \ref{chiral field 1,1 appendix}). Thus, in this limit the GLSM realizes a  K\"ahler geometry on $\M=\{ \mu_2+\xi=0\}/U(1)$. For finite $\beta$ it realizes a generalized K\"{a}hler geometry on $\M$.

The generalization to the multiflavor case is straightforward. For $N_{s}$ pairs of semichiral fields $(\Bbb X_L^i, \Bbb X_R^i)$ with charges $(Q^{i},-Q^{i})$ and action (\ref{many semis chiral gauging}), the Higgs branch corresponds to solutions to
\equ{ \label{quotient manifold opposite charges}
\sum_{i=1}^{N_{s}}Q_{i}\(| X_{L}^{i}|^{2}-|X_{R}^{i}|^{2}\)-\xi=0\,,
}
modulo $U(1)$ gauge transformations, a noncompact manifold of complex dimension $2N_{s}-1$. Since the case of opposite charges is related to the case of equal charges by a simple change of coordinates,  the moduli space is always noncompact, regardless of the charge assignments. Topologically, (\ref{quotient manifold opposite charges}) coincides with the moduli space for $N_{c}=N_s$ fundamental chiral fields with charges $Q_{i}$ and $N_{c}$ antifundamental fields of charges $-Q_{i}$, a class of examples of the theories studied in \cite{Hanany:1997vm}. In fact, in the special limit $\beta_i\to \infty$ these gauge theories coincide, as noted above. However, at finite values of $\beta_i$ these theories are rather different; in particular, for generic values of $\beta_i$ the rigid symmetry of the system  is $U(1)^{N_s-1}\times U(1)_A\times U(1)_V$ (in the absence of twisted masses), as opposed to $SU(N_c)\times SU(N_c)\times U(1)_a\times U(1)_A\times U(1)_V$ in the chiral case. Furthermore, as we shall see in an explicit example, for generic $\beta_i$ the geometry of the Higgs branch has nonzero torsion and is generalized K\"{a}hler rather than K\"{a}hler. 

In the case of chiral and semichiral fields coupled to the standard vector multiplet with an action given by the sum of (\ref{kinetic term chirals}) and (\ref{many semis chiral gauging}) one includes in (\ref{quotient manifold opposite charges}) the usual contribution from chiral fields: $\sum_a q_a|\phi^a|^2$.

We note that promoting the FI parameter $t$ to a twisted chiral field $\chi$  corresponds to constraining \textit{both} field strengths in the SVM to vanish, $i.e.$,  performing a T-duality \cite{Lindstrom:2007sq,Merrell:2007sr}. Thus, the moduli space (\ref{quotient manifold opposite charges}) can also be  thought of as  performing a T-duality of flat space (described in terms of semichiral fields) and taking the slice $\chi=t$.

We may summarize as follows: the space of classical vacua of the theories at hand typically consists of a Coulomb branch and a Higgs branch. In this paper we do not discuss the Coulomb branch further, focusing on the geometry of the Higgs branch. The classical Higgs branch is a noncompact generalized K\"{a}hler manifold whose geometric structure depends on data such as the number of multiplets and their charges, and the vector multiplet they couple to, in addition to a set of continuous parameters $\beta_i$ controlling the deformation from a certain K\"ahler geometry; see Table~\ref{Table target} below.

\subsection{The \textit{type} of the target space}

An important characteristic of a generalized K\"{a}hler structure is its \textit{type} $(k_+,k_-)$, given by 
\equ{
(k_+,k_-)=(\text{dim}_{\Bbb C} \,  \text{ker} (J_+- J_-),\,\text{dim}_{\Bbb C} \,  \text{ker} (J_++ J_-))\,.
}
In terms of $\N=(2,2)$ multiplets, $k_+$ and $k_-$ simply count, respectively, the number of chiral and twisted chiral fields describing the manifold; the number of semichiral fields is given by $\text{dim}_{\Bbb C} \,  \text{coim} [J_+, J_-]=d-k_+-k_-$, with $d$ the complex dimension of the manifold. We now discuss the type of Higgs branch. 

Consider a linear model for $N_s$ semichiral fields and a gauged $U(1)$ isometry. If the isometry is gauged by the SVM, the complex dimension of the resulting Higgs branch is $2 N_s-2$ and is described by $N_s-1$ pairs of semichiral fields; the type is $(0,0)$. If instead the isometry is gauged by the CSVM  the complex dimension of the target is $2N_{s}-1$; in the case $\Bbb F=0$ the type is $(1,0)$\footnote{This is clear from the point of view of the CSVM; one may use the gauge parameters in the SVM to gauge-fix one pair of semichiral fields away, at the expense of introducing the chiral Lagrange multiplier $\Phi$. Alternatively,  in terms of the usual vector multiplet, the gauge parameter allows one to only partially gauge away  a pair of semichiral fields, leaving a complex linear superfield, which is dual to a chiral superfield (Section~\ref{Relation to complex linear superfields}).} and in the case $\tilde{\Bbb F} =0$ the type is $(0,1)$. For GLSMs involving $N_c$ chiral fields and $N_s$ pairs of semichiral fields coupled to the standard vector, the complex dimension is $2 N_s+N_c-1$ and the type is $(N_c-1,0)$. This is summarized in Table~\ref{Table target}.

\begin{table}[h]
\begin{center}
\begin{tabular}{|c|c|c|c|c|c|}\hline
 Vector & Couples to & dim$_\Bbb C$ & Geometry & Type & ${\mathcal A} (R)$\\ \hline
 V & $\Bbb X_L^i,\Bbb X_R^i; \Phi^a $ & $2 N_s+N_c-1$& def. of K\"{a}hler &$(N_c-1,0)$ & $\sum_a q_a$ \\ \hline
 twisted V & $\Bbb X_L^i,\Bbb X_R^i; \chi^{a'} $ & $2 N_s+N_{tc}-1$& def. of K\"{a}hler &$(0,N_{tc}-1)$ & $\sum_{a'} q_{a'}$ \\ \hline
 SVM & $\Bbb X_L^i,\Bbb X_R^i$ & $2 N_s-2$& def. of hyperk\"{a}hler &$(0,0)$&  0 \\ \hline
 CSVM & $\Bbb X_L^i,\Bbb X_R^i$ & $2 N_s-1$& def. of Calabi-Yau& $(1,0)$ or $(0,1)$& 0\\ \hline
\end{tabular}
\end{center}
\caption{Possible couplings of matter fields to various vector multiplets and the geometry of the Higgs branch. Here $i=1,...,N_s,\, a=1,...,N_c\,, a'=1,...,N_{tc}$. In the case of a $U(1)$ gauge group we give the dimension and type (at generic points) of the manifold. For generic values of the parameters $\beta_i$ (or $\alpha_i$) the geometry is generalized K\"{a}hler, but for special values it becomes K\"{a}hler, hyperk\"{a}hler or Calabi-Yau.}\label{Table target}
\end{table}

For higher-rank gauge groups various gaugings are possible. For instance, one may consider a GLSM with $N_s$ pairs of semichiral fields and gauge group $G=U(1)^k \times U(1)^l \times U(1)^m$. Gauging the first factor by the SVM, the second one by the CSVM with $\Bbb F=0$, and the third one by the CSVM with $\tilde{\Bbb F}=0$, leads to a target of complex dimension $2 N_s-2k -l-m$ and type $(l,m)$. One may also introduce charged chiral and twisted chiral multiplets coupled to their corresponding vector multiplets. 

An important aspect of generalized complex geometry is that the type is not necessarily constant over the manifold; it may jump discontinuously on certain loci \cite{Gualtieri:2003dx}.  On these loci---if they exist---a pair of semichiral fields $(\Bbb X_L, \Bbb X_R)$ become either a pair of chiral or a pair of twisted chiral fields (but cannot turn into a chiral and a twisted chiral field \cite{2008arXiv0804.3621G}). The type of the moduli spaces given above and in Table~\ref{Table target} refers to generic points in the manifold, away from possible type-changing loci.  We will discuss this in more detail in Section~\ref{Conifold with torsion} in a specific example where, at least for a specific choice of parameters, there are no type-changing loci. A study of whether type-changing loci exist or not in these models in general is an interesting question which we leave for future work.

\subsection*{\it Comment on RG Flow}

An important question is what are the quantum corrections to the moduli space geometry. In particular, in the undeformed case it is well known that if the R-symmetry anomaly cancels the gauge theory flows in the deep IR to a NLSM on a toric Calabi-Yau. It would be interesting to study the deep IR behavior of the deformed theories, which we do not discuss here. Theorems in supergravity \cite{Maldacena:2000mw,Ivanov:2000fg,Giddings:2001yu} forbid the existence of \textit{compact} Generalized Calabi-Yau manifolds with a non-zero H-flux. The theories we have discussed here avoid this no-go theorem by being noncompact. The conditions for conformal invariance of the NLSM at the quantum level, and the relation to the  generalized Calabi-Yau condition of Hitchin \cite{Hitchin:2004ut}, is discussed in \cite{Halmagyi:2007ft,Grisaru:1997pg,Hull:2010sn}. 

\section{An example: the conifold with torsion}
\label{Conifold with torsion}

Consider two pairs of semichiral fields $(\Bbb X_L^1, \Bbb X_R^1,\Bbb X_L^2, \Bbb X_R^2)$ with charges $(1,-1,1,-1)$, respectively. The action we consider is 
\eqs{  \nonumber
\L=\,&- \frac{1}{2e^{2}}\int d^{4}\theta \(\bar {\tilde{ \Bbb F}} \tilde{\Bbb F} -\bar {\Bbb F} \Bbb F \) +\int d^{4}\theta \sum_{i=1,2}\, \Big[ \Bbb{ \bar X}_L^{i} e^{V_L } \Bbb{ X}_L^{i} + \Bbb{ \bar X}_R^{i} e^{- V_R} \Bbb{ X}_R^{i}+ \beta_i \, ( \Bbb{ X}_L^{i} e^{- i \Bbb V} \Bbb{ X}_R^{i}+c.c. ) \Big] \\  \label{Model conifold opposite charges}
\,&+\(i \int d^{2}\theta \,  \Phi\, \Bbb{ F}+c.c.\)+\(i \, t \int d^{2}\tilde \theta  \, \tilde{\Bbb F}+c.c.\)\,.
}
As discussed, the low-energy dynamics on the Higgs branch is given by a NLSM on 
\equ{
\big| X_{L}^{1}\big|^{2}+\big|X_{L}^{2}\big|^{2}-\big| X_{R}^{1}\big|^{2}-\big|X_{R}^{2}\big|^{2}=\xi\,,
}
modulo $U(1)$ gauge transformations. Topologically, this space coincides with the well-known resolved conifold \cite{Candelas1990246}, which admits a Calabi-Yau metric \cite{PandoZayas:2000sq} with $SU(2)\times SU(2)\times U(1)$ symmetry and can be realized as a GLSM for chiral fields with charges $(1,1,-1,-1)$ \cite{Klebanov:1998hh}. The realization by the new GLSM  (\ref{Model conifold opposite charges}) leads to a two-parameter family of generalized K\"{a}hler structures on the space. For generic $\beta_{1,2}$ the flavor symmetry of the system is broken to $U(1)$, which is enhanced to $SU(2)$ in the special case $\beta_1=\beta_2$. To obtain the generalized K\"{a}hler potential describing this space we integrate out the CSVM in superspace. For simplicity we will set now $\beta_1=\beta_2\equiv \beta$ and $t=0$. In the limit $e\to \infty$ the equations of motion read:
\eqsn{
\Bbb{ \bar X}_L^{i} e^{V_L } \Bbb{ X}_L^{i} - \Bbb{ \bar X}_R^{i} e^{- V_R} \Bbb{ X}_R^{i}=\,&0\,,\\
\Bbb{ \bar X}_L^{i} e^{V_L } \Bbb{ X}_L^{i}+ \Bbb{ \bar X}_R^{i} e^{- V_R} \Bbb{ X}_R^{i}+ \beta \, (\Bbb{ X}_L^{i} e^{- i \Bbb V} \Bbb{ X}_R^{i}+c.c. )=\,&2\, \text{Re}\,\Phi\,,\\
 \frac{i \beta}{2} \, ( \Bbb{ X}_L^{i} e^{- i \Bbb V} \Bbb{ X}_R^{i}-c.c. )=\,&-\text{Im}\, \Phi\,.
}
Solving these equations for the vector multiplet and evaluating the action (\ref{Model conifold opposite charges}) gives the generalized K\"{a}hler potential. In terms of the gauge-invariant combinations
\equ{
\Bbb X_L = \frac{\Bbb X_L^1}{\Bbb X_L^2}\,, \qquad \Bbb X_R = \frac{\Bbb X_R^1}{\Bbb X_R^2}\,,
}
(which are good target coordinates in the patch $\Bbb X_L^2,\Bbb X_R^2\neq 0$) and the Lagrange multiplier $\Phi$, the generalized potential reads:
\equ{ \label{KUV conifold}
K_{UV}=-\frac{1}{2} \Phi \log F -\frac{1}{2} \bar \Phi \log \bar F\,,
}
where $F=F(\Phi, \bar \Phi; \Bbb X_{L}, \bar{\Bbb X}_{L},\Bbb X_{R}, \bar{\Bbb X}_{R})$ is given by 
\equn{
F= \frac{2\beta^{2}\Phi |1+\Bbb X_{L}\Bbb X_{R}|^{2} -(1+|\Bbb X_{L}|^{2})(1+|\Bbb X_{R}|^{2})(\Phi-\bar \Phi) -\sqrt{(1+|\Bbb X_{L}|^{2})(1+|\Bbb X_{R}|^{2})}\, G}{4 \beta (1+\Bbb X_{L}\Bbb X_{R})(-(1+|\Bbb X_{L}|^{2})(1+|\Bbb X_{R}|^{2})+\beta^{2}|1+\Bbb X_{L}\Bbb X_{R}|^{2} )}
}
and
\equ{G=\sqrt{(1+|\Bbb X_{L}|^{2})(1+|\Bbb X_{R}|^{2})(\Phi-\bar \Phi)^{2}+4 \beta^{2} |\Phi|^{2}|1+\Bbb X_{L}\Bbb X_{R}|^{2}}\,.}

The subscript `UV' in (\ref{KUV conifold}) is a reminder that although this is a description at low energies compared to the scale set by the gauge coupling constant $e$, it is still at high energies from the point of view of the NLSM; this NLSM is not conformal and will flow under RG. Using the formulas in Appendix~\ref{Metric and B-field} one can compute $(g,H,J_\pm)$ explicitly from (\ref{KUV conifold}), but the full expressions are lengthy and not particularly enlightening. We emphasize that the H-field is non-zero:
\eqsn{
H_{UV}= \frac{ d\phi \wedge dX_L \wedge d\bar X_L}{\beta^2\(1+|X_L|^2\)^2}-\frac{ d\phi \wedge dX_R \wedge d\bar X_R}{\beta^2\(1+|X_R|^2\)^2}+c.c.+ \mathcal O (\beta^{-3})\,.
}
As expected, in the special limit $\beta \to \infty$ the H-field vanishes, both complex structures $J_\pm$ become the canonical ones, and the geometry becomes K\"{a}hler. For finite $\beta$ the space is generalized K\"{a}hler, as ensured by $\N=(2,2)$ SUSY.  

As mentioned, the generalized potential (\ref{KUV conifold}) does not describe a conformal model and will flow nontrivially under RG. In the case $\beta \to \infty$ the endpoint of the flow is a NLSM on a Calabi-Yau; it would be interesting to study possible RG fixed points for finite $\beta$. One approach to investigate this question would be to study solutions to the generalized Monge-Amp\`{e}re equation \cite{Hull:2010sn}. The specific case of one pair of semichiral fields and one chiral field, as here, is of particular interest for supergravity and was studied in \cite{Halmagyi:2007ft}.  We leave this for future work.

One may also consider a GLSM for a single pair of semichiral fields and two chiral fields $(\Bbb X_L, \Bbb X_R;\Phi_1,\Phi_2)$ with $U(1)$ charges $(1,-1;1,-1)$, respectively, gauged by the standard vector multiplet $V$. This model realizes a family of generalized K\"{a}hler structures on the conifold with type $(1,0)$ and $U(1)^2$ isometry; this model is discussed in more detail in \cite{Benini:2015isa}. 

\subsection*{\it Type-change analysis}

In the current example, at generic points on the manifold the type is $(k_+,k_-)=(1,0)$. To investigate the possibility of type change we study the eigenvalues of $\tfrac12(J_+-J_-)$  and $\tfrac12(J_++J_-)$. These are given by $\{0,\pm i \lambda\}$ and $\{\pm i,\pm i \tilde \lambda\}$, respectively, with $\lambda, \tilde \lambda$ given in (\ref{eigen chiral appendix},\ref{eigen t chiral appendix}). Interestingly, for the potential  (\ref{KUV conifold}) we find that the eigenvalues are constant over the manifold:
\equ{\label{eigen conifold}
\lambda= \frac{1}{\beta}\,, \qquad \tilde \lambda=\sqrt{1-\frac{1}{\beta^2}} \,.
}
Thus, there are no type-change loci in this coordinate  patch. For $\beta\to\infty$ the manifold is described at every point by chiral fields. The eigenvalues (\ref{eigen conifold}) are unmodified by turning on an FI parameter $t\neq0$, but are no longer constant when $\beta_1 \neq \beta_2$. We leave a careful study of the general case for future work.

\section{Discussion and Outlook}
\label{Discussion}

In this paper we have described $\N=(2,2)$ GLSMs that arise from coupling chiral, twisted chiral, and semichiral fields to various $\N=(2,2)$ vector multiplets. These include the standard vector and twisted vector multiplets, the SVM, and a new kind of gauge multiplet which we call the CSVM. The GLSMs depend on a set of continuous parameters $\beta_i$ that control, among other couplings, a gauged $b$-field term. For special values of the parameters these become standard GLSMs realizing K\"{a}hler manifolds, but for generic values they are non-K\"{a}hler deformations. The deformation preserves R-symmetry at the quantum level, but typically breaks some rigid isometries. 

An interesting class of models arises from coupling semichiral fields to the CSVM. These are deformations of GLSMs for chiral fields containing an equal number of fundamental and antifundamental fields, realizing NLSMs on noncompact Calabi-Yau manifolds. As a specific example we have discussed a GLSM realizing a two-parameter family of generalized K\"{a}hler structures on the conifold, with only a $U(1)$ isometry for generic value of the parameters. In the case in which the isometry is enhanced to an $SU(2)$ we have shown that there are no type-changing loci. It would be interesting to study whether type-changing loci are present in these models generically. This is relevant \cite{Kapustin:2003sg,Kapustin:2004gv} to the study of nonperturbative corrections in the topologically twisted version of  NLSMs with H-flux \cite{Kapustin:2004gv} (see also \cite{Bredthauer:2006hf,Zucchini:2006ii,Chuang:2006vt}). In the present context this issue is discussed in more detail in \cite{Benini:2015isa}.  

Another interesting class of theories arises from coupling to the SVM; as discussed these lead to deformations of hyperk\"{a}hler manifolds and will be analyzed in more detail elsewhere.

There are various open questions. Regarding the new CSVM our discussion has been restricted to $U(1)^k$ gauge theories, but it may be interesting to study possible constraints on the nonabelian SVM as well. Even in the Abelian case there are important open questions, most importantly the deep IR behavior of these NLSMs. One may also study the T-dual description of these gauge theories including nonperturbative aspects, along the lines of \cite{Hori:2000kt,Morrison:1995yh}. An understanding of this could shed light on mirror symmetry in a generalized context.  Finally, it would be interesting to formulate this method for constructing generalized K\"ahler manifolds as a quotient in the context of generalized K\"{a}hler geometry.

One may also consider GLSMs with torsion without introducing semichiral fields, by coupling chiral and twisted chiral fields to the Large Vector Multiplet (LVM); these can realize compact manifolds and will be studied elsewhere. 

\section*{\centering Acknowledgements}
We would like to especially thank Francesco Benini and Dharmesh Jain for many valuable discussions and collaboration on related problems. We also acknowledge valuable discussions with Nikolay Bobev, Ulf Lindstr\"{o}m, Daniel Park, Stefan Vandoren, and Rikard von Unge. 

P.M.C. is supported by the Netherlands Organization for Scientific Research (NWO) under the VICI Grant 680-47-603. This work is part of the D-ITP consortium, a program of the NWO that is funded by the Dutch Ministry of Education, Culture and Science (OCW). MR acknowledges NSF 
Grant No. PHY-1316617.  P.M.C. would like to thank the ``2014 Summer Simons Workshop in Mathematics and Physics'' at Stony Brook, during which part of this work was done.

\appendix

\section{Target space geometry}\label{Metric and B-field}
Here we give relevant formulas for computing the metric, $b$-field, and complex structures from the generalized K\"{a}hler potential (for a review and details see \cite{Lindstrom:2005zr}). The action for a $\N=(2,2)$ NLSM  is given by 
\equ{
\L=\int d^4\theta \,K(\Phi,\bar \Phi;\chi,\bar \chi; \Bbb X_L,\bar{ \Bbb X}_L,\Bbb X_R, \bar{ \Bbb X}_R )\,,
}
where $\Phi, \chi, \Bbb X_L,\Bbb X_R$ denote chiral, twisted chiral, and left and right semichiral fields, respectively, and $d^4 \theta$ is the Grassmann measure. The function $K$ is known as the generalized K\"{a}hler potential and apart from obeying mild conditions for the metric to be positive definite, it is otherwise arbitrary. The metric, $b$-field, and complex structures are completely determined by this function. Defining $E=\frac12(g+B)$ one has
\eqs{
E_{LL} =\,& C_{LL}K_{LR}^{-1}J_sK_{RL} \cr
E_{LR} =\,& J_sK_{LR}J_s + C_{LL}K_{LR}^{-1}C_{RR} \cr
E_{Lc} =\,& K_{Lc} + J_s K_{Lc} J_c + C_{LL}K_{LR}^{-1}C_{Rc}\cr
E_{Lt}=\,& -K_{Lt} - J_s K_{Lt} J_t + C_{LL}K_{LR}^{-1}A_{Rt}\cr
E_{RL} =\,& -K_{RL}J_s K_{LR}^{-1} J_s K_{RL}\cr
E_{RR} =\,& -K_{RL}J_s K_{LR}^{-1} C_{RR}\cr
E_{Rc} =\,& K_{Rc} - K_{RL}J_s K_{LR}^{-1} C_{Rc}\cr
E_{Rt} =\,& -K_{Rt} - K_{RL}J_s K_{LR}^{-1} A_{Rt}\\
E_{cL} =\,& C_{cL}K_{LR}^{-1}J_s K_{RL}\cr
E_{cR}=\,& J_c K_{cR} J_s + C_{cL}K_{LR}^{-1}C_{RR}\cr
E_{cc} =\,& K_{cc}+J_c K_{cc} J_c + C_{cL}K_{LR}^{-1}C_{Rc}\cr
E_{ct} =\,& -K_{ct}-J_c K_{ct}J_t + C_{cL}K_{LR}^{-1}A_{Rt}\cr
E_{tL} =\,& C_{tL}K_{LR}^{-1}J_s K_{RL}\cr
E_{tR} =\,& J_t K_{tR} J_s + C_{tL}K_{LR}^{-1}C_{RR}\cr
E_{tc} =\,& K_{tc} + J_t K_{tc} J_c + C_{tL}K_{LR}^{1}C_{Rc}\cr
E_{tt} =\,& -K_{tt} - J_t K_{tt} J_t + C_{tL} K_{LR}^{-1} A_{Rt}\nonumber
}
where $A$, $C$, and $K$ are matrices whose entries are second derivatives of the generalized potential and $c, t, s$ denote chiral, twisted chiral and semichiral directions, respectively. For instance
\equ{\label{definition KLR}
K_{LR}\equiv\pmat{\frac{\partial^2 K}{\partial X_L \partial X_R} & \frac{\partial^2 K}{\partial X_L \partial \bar{X}_R} \\ \frac{\partial^2 K}{\partial \bar{X}_L \partial X_R} & \frac{\partial^2 K}{\partial \bar{X}_L \partial \bar{X}_R}},
}
and similarly for $K_{Rc}$, etc, and
\equ{
A\equiv \pmat{ 2i K&  0 \\ 0  & -2 i K}\,,\qquad C \equiv\pmat{   0 & 2i K \\  -2 i K & 0 }\,,
}
where $K$ is the matrix of second derivatives defined above.

The complex structures read \cite{Lindstrom:2005zr,Bogaerts:1999jc}
\eqs{\nonumber
J_{+}=&
\left(\begin{array}{cccc}
J_s &0&0&0\cr
K_{RL}^{-1}C_{LL} &  K_{RL}^{-1}J_s K_{LR} &
K_{RL}^{-1}C_{Lc}
& K_{RL}^{-1}C_{Lt}\cr
0&0&J_c&0\cr
0&0&0&J_t\end{array}\right)\,,~
\\
J_{-}=&\left(\begin{array}{cccc}
K_{LR}^{-1}J_s K_{RL} & K_{LR}^{-1}C_{RR} &
K_{LR}^{-1}C_{Rc}& K_{LR}^{-1}A_{Rt}\cr
0& J_s&0&0\cr
0&0&J_c&0\cr
0&0&0& -J_t\end{array}\right)\,,
\label{complex structures general}
}
where $J_{c,t,s}$ are canonical complex structures of the form $diag(i,-i)$ of the appropriate dimension. 

\subsection*{Flat Space}
\label{Flat Space Appendix}

Setting $c=0$ in the flat-space Lagrangian (\ref{matter action general}) (and making a phase redefinition to make $d$ real), in the basis $X^{\mu}=(X_L,\bar X_L, X_R, \bar X_R)$ one has
\equ{ \label{J's flat space}
J_+=\left(
\begin{array}{cccc}
 i & 0 & 0 & 0 \\
 0 & -i & 0 & 0 \\
 0 & \frac{2 i a}{d} & i & 0 \\
 -\frac{2 i a}{d} & 0 & 0 & -i \\
\end{array}
\right)\,, \qquad J_-=\left(
\begin{array}{cccc}
 i & 0 & 0 & \frac{2 i b}{d} \\
 0 & -i & -\frac{2 i b}{d} & 0 \\
 0 & 0 & i & 0 \\
 0 & 0 & 0 & -i \\
\end{array}
\right)\,,
}
and
\equ{g=4
\left(
\begin{array}{cccc}
 0 &  a & \frac{ a b}{d} & 0 \\
  a & 0 & 0 & \frac{ a b}{d} \\
 \frac{ a b}{d} & 0 & 0 &  b \\
 0 & \frac{ a b}{d} &  b & 0 \\
\end{array}
\right)\,,\qquad  b= \tfrac{2\(2ab-d^2\)}{d} \left(
\begin{array}{cccc}
 0 & 0 & 1 & 0 \\
 0 & 0 & 0 & 1 \\
  -1 & 0 & 0 & 0 \\
 0 &  -1 & 0 & 0 \\
\end{array}
\right)\,.
\label{g and b flat space}
}

\subsection*{Type-change}
\label{Type-change}

To study the phenomenon of type-change one may compute the eigenvalues of $J_+\pm J_-$. For concreteness we consider here a generalized potential $K$ that depends on a single chiral field and a pair of semichiral fields. Thus, at generic points in the manifold the type is $(k_+,k_-)=(1,0)$. From (\ref{complex structures general}) it is easy to see that the eigenvalues of $\tfrac12(J_+-J_-)$ are $\{0,\pm i \lambda\}$, each of multiplicity two, and 
\equ{\label{eigen chiral appendix}
\lambda= \sqrt{
\frac{|K_{\bar l r}|^2-K_{r\bar r} K_{l\bar l}}{|K_{\bar l r}|^2-|K_{l r}|^2}
}\,.
}
The eigenvalues of $\tfrac12(J_+ + J_-)$ are $\{\pm i,\pm i \tilde \lambda\}$, each of multiplicity two, and 
\equ{ \label{eigen t chiral appendix}
\tilde \lambda= \sqrt{
\frac{|K_{l r}|^2-K_{r\bar r} K_{l\bar l}}{|K_{l r}|^2-|K_{\bar l r}|^2}
}\,.
}
We note the relation $\lambda^2+\tilde \lambda^2=1$. On the locus $\lambda=0$ the type jumps to $(k_+,k_-)=(3,0)$, where the manifold is locally described by three chiral fields, and on the locus $\tilde \lambda=0$ the type jumps to $(k_+,k_-)=(1,2)$, where it is locally described by one chiral field and two twisted chiral fields. Whether these loci exist or not depends on the specific potential $K$. 

\section{Gauging and $\N=(2,2)$ Vector Multiplets}
\label{Vector Multiplets}

In this and the next appendix we review some basic elements of standard and novel $\N=(2,2)$ vector mutiplets. We mostly follow \cite{Lindstrom:2007vc, Lindstrom:2008hx,Lindstrom:2005zr}.

\subsection*{\it Vector and Twisted Vector Multiplet}

The standard vector multiplet $V$ transforms under gauge transformations as
\equ{
e^V \to e^{i \bar \Lambda} \, e^V \, e^{- i \Lambda}\,.
}
It is convenient to introduce gauge-covariant superderivatives $\nabla_\pm, \bar \nabla_\pm$. In the case of the standard vector multiplet these satisfy
\equ{
\{\nabla_\pm, \bar{\nabla}_\pm\}= 2i \,  \mathcal D_{\pm\pm} \,, \qquad \Sigma = i \,  \{\bar \nabla_{+},\nabla_{-}\}\,, 
}
where $\mathcal D_{\pm\pm} $ is the spacetime gauge-covariant derivative, $\Sigma$ is a twisted chiral field strength, and all other anticommutators vanish.

Gauge-covariant derivatives can be expressed in terms of the superderivatives  $\Bbb D_\pm, \bar{\Bbb D}_\pm$. There are different representations, depending on the matter fields they act on. In chiral representation, for instance, all objects transform with a chiral gauge parameter; if a chiral field transform as
\equ{
\Phi\to e^{i \Lambda} \, \Phi\,, \qquad  \bar \Phi\to \bar \Phi \, e^{-i \bar \Lambda}\,,
}
one defines the fields $\hat \Phi=\Phi$ and $\bar{\hat \Phi}=\bar \Phi e^V$, with gauge transformations $\hat \Phi\to e^{i \Lambda}\hat \Phi$ and $\bar{\hat \Phi} \to \bar{\hat \Phi} e^{- i \Lambda}$. These fields are chiral (antichiral) with respect to the gauge-covariant derivatives: $\bar{ \nabla}_\pm \hat \Phi= \nabla_\pm \bar{\hat  \Phi}=0$, where
\equ{ \label{chiral rep chiral field}
\bar \nabla_{\pm}=\bar{ \Bbb D}_{\pm}\,, \qquad \nabla_{\pm}= e^{-V} \Bbb{D}_\pm e^V\,,}
and transform as 
\equ{\bar\nabla_{\pm} \to  e^{i \Lambda} \bar\nabla_{\pm}  e^{-i \Lambda}\,,\qquad\nabla_{\pm} \to  e^{i \Lambda} \nabla_{\pm}  e^{-i \Lambda}\,.} 
In the Abelian case $\Sigma=i \, \{\bar \nabla_{+},\nabla_{-}\}=i \bar{ \Bbb D}_+ \Bbb{D}_- V$.

Gauge-invariant actions for matter fields are written as
\equ{
\int d^4 \theta \, \bar \Phi e^V \Phi = \int d^4 \theta \, \bar{\hat \Phi} \hat \Phi= \(\nabla_+ \nabla_- \bar{\nabla}_+ \bar{\nabla}_-\, (\bar{\hat \Phi} \hat \Phi)\)\Big|\,,
}
where $|$ means setting  $\theta^\pm=\bar \theta^\pm=0$.

\vspace{1cm}

The twisted vector multiplet $\tilde V$ transforms under gauge transformations as
\equ{
e^{\tilde V} \to e^{i \bar{\tilde \Lambda}} \, e^{\tilde V} \, e^{- i \tilde \Lambda}\,.
}
Gauge-covariant derivatives satisfy
 \equ{
\{\nabla_\pm, \bar{\nabla}_\pm\}= 2i \,  \mathcal D_{\pm\pm} \,, \qquad \Theta = i \,  \{\bar \nabla_{+},\bar{\nabla}_{-}\}\,, 
}
where $\Theta$ is a chiral field strength and all other anticommutators vanish. As above, one can construct gauge-covariant derivatives in the corresponding representation.

\subsection*{\it Semichiral Vector Multiplet}

The semichiral vector multiplet  \cite{Lindstrom:2008hx} consists of  $(V_L,V_R, \Bbb V, \Bbb{\tilde V})$, where $V_{L,R}$ are real and $\Bbb V, \tilde{\Bbb V}$ are complex. These transform as
\eqsn{
e^{V_L}\to & e^{i \bar \Lambda_L} \, e^{V_L}  e^{- i \Lambda_L}\\
e^{V_R}\to & e^{i \bar \Lambda_R} \, e^{V_R}  e^{- i \Lambda_R}\\
e^{i \Bbb V}\to & e^{i \Lambda_L} \, e^{i \Bbb V}  e^{- i \Lambda_R}\\
e^{i \tilde{\Bbb V}}\to &  e^{i \Lambda_L} \, e^{i \tilde{\Bbb V}}  e^{- i \bar{ \Lambda}_R}\,,
}
and satisfy the gauge-covariant constraints
\equ{
e^{i \bar{ \tilde{\Bbb V} }}=e^{V_L}e^{i \Bbb V}\,, \qquad e^{i \bar{\Bbb V}}= e^{V_L}e^{i {\Bbb V}}e^{-V_R}\,.
}
Gauge-covariant derivatives satisfy: 
\equ{
\{\nabla_\pm, \bar{\nabla}_\pm\}= 2i \,  \mathcal D_{\pm\pm} \,, \qquad \tilde{\Bbb F }= i \, \{\bar \nabla_+, \nabla_-\}\,, \qquad   \Bbb F=i \, \{\bar \nabla_+, \bar \nabla_-\}\,,
}
where $\tilde{\Bbb F }$ and $\Bbb F$ are twisted chiral and chiral, respectively, and all other anticommutators vanish.

In left semichiral representation all objects transform with a left semichiral parameter $\Lambda_{L}$. One defines 
\eqs{\nonumber
\Hat{ \Bbb X}_L=\,&\Bbb{X}_{L}\,\\ \nonumber
\bar{\Hat{ \Bbb X}}_L=\,&\bar{\Bbb{X}}_{L}\,e^{V_{L}}\,\\
\Hat{ \Bbb X}_R =\, & e^{i\Bbb V}  \Bbb X_R\, \\  \nonumber
\bar{ \Hat{\Bbb X}}_R =\,& \bar{\Bbb X}_R \,e^{- i \tilde{\Bbb V} }\,;
}
all these transform with a left semichiral parameter under gauge transformations. Gauge-covariant derivatives are given by
\eqs{\nonumber
\bar{\nabla}_{+}=\, & \bar{\Bbb{D}}_{+}\\ \nonumber
\nabla_{+}=\,& e^{-V_{L}}\Bbb D_{+} e^{V_{L}}\\
\bar{\nabla}_{-}=\, & e ^{i \Bbb V} \bar{\Bbb{D}}_{-}e^{- i \Bbb V}\\ \nonumber
\nabla_{-}=\, & e ^{i \tilde{\Bbb V}} \Bbb{D}_{-}e^{- i\tilde{ \Bbb V}}\,.
}
Of course, one could also consider a right semichiral representation, etc.

The hatted fields are semichiral with respect to gauge-covariant derivatives: $\bar{\nabla}_+ \hat{\Bbb X}_L=\bar{\nabla}_- \hat{\Bbb X}_R=\nabla_+ \bar{\hat{\Bbb X}}_L=\nabla_- \bar{\hat{\Bbb X}}_R=0$. Gauge-invariant actions for matter fields are written as, $e.g.$,
\equ{ \label{lagrangian semis coupled to SVM covariant approach opp charges}
\int d^4\theta \, \( \bar{\hat { \Bbb X}}_L \hat{\Bbb X}_L + \bar{ \hat{\Bbb X}}_R \hat{\Bbb X}_R +\beta ( \hat{\Bbb X}_L \hat{\Bbb X}_R + \bar{ \hat{\Bbb X}}_R \bar{\hat{\Bbb X}}_L)\)\,.
}

In the Abelian case, $\tilde{\Bbb F}\equiv \bar{\Bbb D}_{+}\Bbb D_{-} \tilde{\Bbb V}$ and  $\Bbb F\equiv\bar{\Bbb D}_{+}\bar{\Bbb D}_{-} \Bbb V$ and 
\equ{\label{relations SVM appendix}
-\frac{1}{2}V'\equiv \text{Re} \, \Bbb{\tilde V}=\text{Re}\, \Bbb V \,,\qquad \text{Im}\, (\Bbb{\tilde V} -  \Bbb{V})=V_{R}\,,\qquad \text{Im}\,( \Bbb{\tilde V} +  \Bbb{V})=V_{L}\,. 
}
Thus, one may write 
\equ{\label{V and V tilde}
\Bbb V=\frac{1}{2}(-V'+i (V_L-V_R))\,, \qquad \tilde{ \Bbb V} =\frac{1}{2}(-V'+i (V_L+V_R))\,,
}
where $V'$ transforms under gauge transformations as $\delta  V' =(\Lambda_R + \bar  \Lambda_R- \Lambda_L -\bar  \Lambda_L)$. In terms of these, the FI terms (\ref{FI terms SVM}) read
\equ{\label{FIappB}
\L_{FI}=-\int d^4\theta \( \text{Re}(s+t) V_L-\text{Re}(s-t) V_R-\text{Im}(s+t)V'\)\,.
}
The combination $\text{Im}(s-t)$ is a theta parameter and couples to the field strength $F_{01}$, which we should be added to (\ref{FIappB}).

\section{Reduction to components}

\subsection{Reduction to $\N=(1,1)$}
\label{Reduction to $(1,1)$}

As reviewed in Appendix~\ref{Vector Multiplets} in the presence of gauge fields it is convenient to work in the covariant approach. In what follows, we always work with gauge-covariant derivatives $\nabla_\pm, \bar{\nabla}_\pm$ and the appropriate matter fields satisfying gauge-covariant SUSY constraints (in this section we drop the `hat' symbol in the matter fields to avoid cluttering the notation).

To reduce to $\mathcal{N}=(1,1)$ superspace, one writes 
\equ{\label{2,2 derivatives to 1,1}
\nabla_{\pm}=\frac{1}{2}\left(  \mathcal{D}_{\pm} -i \mathcal{Q}_{\pm}\right), \qquad  \bar{\nabla}_{\pm}=\frac{1}{2}\left(  \mathcal{D}_{\pm} + i \mathcal{Q}_{\pm}\right)\,,
}
where $\mathcal{D}_{\pm}$ are $\mathcal{N}=(1,1)$ gauge-covariant fermionic derivatives and $\mathcal{Q}_{\pm}$ are the gauge-covariant generators of the non-manifest supersymmetry.  The fermionic derivatives satisfy
\equ{\label{commutation D's 1,1}
\{  \mathcal{D}_{\pm} ,  \mathcal{D}_{\pm} \}= i  \mathcal{D}_{\pm \pm}\,,\qquad  \{\D_+,\D_-\}=f\,,
}
where $\mathcal{D}_{\pm \pm}$ is the gauge-covariant space derivative and $f$ is the $\N=(1,1)$ field strength.

D-terms are reduced from $\N=(2,2)$ superspace by writing
\equ{
\int d^{4} \theta K=\int  \D_{+}\D_{-} \(\Q_{+}\Q_{-}K\)\Big|\,,
}
where $|$ means setting $\bar \theta^\pm-\theta^\pm=0$ and F-terms reduce trivially by 
\equ{
\int d^{2} \, \theta \, \mathcal W(\Phi)+\int d^{2} \tilde \theta \, \widetilde{\mathcal W}(\chi)=\int  \D_{+}\D_{-}\( \mathcal{W}(\phi)+\widetilde{\mathcal W}(\chi)\) \,.
}

\subsection*{\it Standard vector multiplets}

The usual $\N=(2,2)$ vector multiplet $V$ transforms under gauge transformations as  $\delta_g V= i (\bar \Lambda-\Lambda)$. The gauge-invariant field strength is defined by $\Sigma= i \{\bar \nabla_+,\nabla_-\}$ and is twisted chiral. The reduction to $\N=(1,1)$ is given by
\equ{
\Sigma | = \frac{1}{2}( \sigma+i f)\,,
}
where $ \sigma$ is a real bosonic superfield and $f$ is defined in (\ref{commutation D's 1,1}). The non-manifest SUSY acts by 
\eqs{ \label{commutation relations Q and D for V}
\{\Q_+,\mathcal{D}_-\} =\,  -   \sigma,\, \qquad 
\{\Q_-,\mathcal{D}_+ \} =\,    \sigma\,,}
and obeys the algebra
\equ{
 \{\Q_+,\Q_- \} =\,  f\,.
}

Similarly, for the twisted vector multiplet $\tilde V$ transforms with a twisted chiral gauge parameter: $\delta_g \tilde V= i (\bar{\tilde \Lambda}-\tilde{\Lambda})$.  The gauge-invariant field strength is defined by $\Theta= i \{\bar \nabla_+,\bar \nabla_-\}$ and is chiral. The reduction to  $\N=(1,1)$ is given by
\equ{
\Theta | = \frac{1}{2}( \sigma'+i f)\,,
}
where $ \sigma'$ is a bosonic superfield, where now:
\eqs{   \label{commutation relations Q and D for Vt}
\{\Q_+,\mathcal{D}_-\} =\,  -  \sigma',\, \qquad 
\{\Q_-,\mathcal{D}_+\} =\,  -  \sigma'\,, \qquad
 \{\Q_+,\Q_- \} =\,  -f\,.
 }

\subsection*{\it Chiral and twisted chiral matter fields}

 Chiral fields $\Phi$ and twisted chiral fields $\chi$ are defined by
\equ{\label{constraints 2,2 fields covariant}
 \bar \nabla_{+}\Phi= \bar \nabla_{-}\Phi=\bar \nabla_{+}\chi= \nabla_{-}\chi=0\,.
}
In $\N=(1,1)$ superspace, chiral and twisted chiral fields both reduce to an unconstrained complex bosonic multiplet; what distinguishes them is their transformation under the non-manifest supersymmetry generated by $\Q_\pm$, which follows from (\ref{constraints 2,2 fields covariant}) and (\ref{2,2 derivatives to 1,1}):
\equ{
\Q_\pm \phi = i \D_\pm \phi \,, \qquad \Q_\pm \chi = \pm i \D_\pm \chi\,, 
}
where $\phi, \chi$ denote the obvious projections.

Using the formulae above, the  kinetic action for a chiral field coupled to the usual vector reads
\equ{\label{chiral field 1,1 appendix}
\int d^4 \theta \, \bar \Phi \Phi =2\int \D_+\D_-\(\D_+ \phi \, \D_- \bar \phi+\D_+\bar \phi \, \D_- \phi -  i \bar \phi \, d\, \phi\)\,,  
}
and for a twisted chiral field coupled to the twisted vector
\equ{ \label{twisted chiral field 1,1 appendix}
-\int d^4 \theta \, \bar \chi \chi = 2 \int \D_+\D_-\(\D_+ \chi \, \D_- \bar \chi+\D_+\bar \chi \, \D_- \chi -  i \bar \chi \, d'\, \chi\)\,.  
}

\subsection*{\it SVM}

There are two gauge-invariant field strengths in the SVM: $\Bbb F = i \{\bar \nabla_+, \bar \nabla_-\}$ and $\tilde{\Bbb F } =i \{\bar \nabla_+, \nabla_-\}$. In terms of $\N=(1,1)$ fields, the SVM consists of three real bosonic superfields\footnote{These fields were previously denoted $d_I$ in references  \cite{Lindstrom:2007vc,Lindstrom:2008hx}.} $\sigma_I$ and the vector multiplet $f$. These are defined by \cite{Lindstrom:2007vc}
\eqs{ \nonumber
\sigma_1=\left( \Bbb F+\bar{\Bbb F} \right) \big|\,, \quad  \sigma_2=&\left( \tilde{\Bbb F}+\bar{\tilde{\Bbb F}} \right)\big|\,, \quad   \sigma_3=i \left(  \Bbb F -\bar{\Bbb F} -\tilde{\Bbb F}+\bar{\tilde{\Bbb F}} \right)\big|\,,   \\ f=&-i \left(  \Bbb F -\bar{\Bbb F} +\tilde{\Bbb F}-\bar{\tilde{\Bbb F}} \right)\big|\,.
}
Equivalently, 
\equ{ \label{definition 1,1 components SVM}
\Bbb F \big|  = \frac{1}{2} \left( \sigma_1+\frac{i}{2} (f-\sigma_3) \right), \qquad \Bbb{\tilde F} \big|  = \frac{1}{2} \left( \sigma_2+\frac{i}{2} (f+\sigma_3) \right)\,.
}
The non-manifest SUSY obeys
\eqs{ \nonumber
\{\Q_+,\mathcal{D}_-\} =\, & - (\sigma_1+\sigma_2),\, \\  \label{commutation relations Q and D for SVM}
\{\mathcal{D}_+,\Q_- \} =\, & - (\sigma_1-\sigma_2)\,, \\  \nonumber
 \{\Q_+,\Q_- \} =\, &  \sigma_3\,.
}
 The reduction of the kinetic action (\ref{action SVM}) gives (in the Abelian case) \cite{Lindstrom:2008hx}
\equ{\label{kinetic action SVM 1,1}
\L_{SVM}=\frac{1}{2e^{2}}\int \D_+\D_-\(\frac{1}{2} \D_{+}f \D_{-}f+ \D_{+}\sigma_{1} \D_{-}\sigma_{1}+\D_{+}\sigma_{2} \D_{-}\sigma_{2}+\frac{1}{2} \D_{+}\sigma_{3} \D_{-}\sigma_{3} \)\,.
}
\subsection*{\it CSVM and twisted masses}
The reduction of the twisted mass constraint (\ref{twisted mass chiral}) gives:
\eqs{ \label{Lphi 1,1}
\L_{\phi}=\,&\frac{i}{2} \int \D_+ \D_-\, \Big[ \phi\(\sigma_1-2\,\text{Re}\, M+ i \(f-\sigma_3- 4\, \text{Im}\,M\)\)\Big]+c.c.\,.
}
Thus, in $\N=(1,1)$ language the CSVM with $\Bbb F=M$ corresponds to setting $\sigma_1=2\, \text{Re}\, M$ and $\sigma_3=f-4 \,\text{Im}\,M$, while constraining  $\tilde{\Bbb F}=\tilde M$ corresponds to setting $\sigma_2=2\, \text{Re}\, \tilde M$ and $\sigma_3=-f+4\, \text{Im} \,\tilde M$ (this also follows directly from \ref{definition 1,1 components SVM}).

\subsection*{\it Semichiral fields}

Left and right semichiral field are defined by 
\equ{\label{constraints 2,2 fields covariant semis}
\bar \nabla_{+}\Bbb X_L=\bar \nabla_{-}\Bbb X_R=0\,.
}
From (\ref{2,2 derivatives to 1,1}), it follows that
\equ{
\Q_+  X_L = i \D_+  X_L\,, \qquad \Q_- X_R = i \D_-  X_R\,.
}
The action of $\Q_-\,(\Q_+)$ on $X_L\,(X_R)$ is undetermined and is simply defined as an independent fermionic multiplet $\Psi_-\,(\Psi_+)$. Thus, in $\N=(1,1)$ language semichiral multiplets consist of a bosonic and fermionic multiplet: 
\eqs{\nonumber 
X_L =& \Bbb X_L\big|\,, \qquad    \Psi_{-} = \Q_- \Bbb X_L \big|\,, \qquad \bar X_L = \Bbb X_L\big|\,, \qquad  \bar \Psi_{-}= \Q_- \bar{\Bbb X}_L \big| \,, \\ \label{1,1 components XL XR}
X_R =& \Bbb X_R\big|\,, \qquad    \Psi_{+} = \Q_+ \Bbb X_R \big|\,, \qquad \bar X_R = \Bbb X_R\big|\,, \qquad  \bar \Psi_{+}= \Q_+ \bar{\Bbb X}_R \big| \,.  
}

Actions such as (\ref{SVM gauged action opposite charge}) are reduced as
\eqs{\nonumber
\L_{matter}=\int \D_+\D_- \Big[\Q_+ \Q_- \( \Bbb{ \bar{ X}}_L  \Bbb{ X}_L + \Bbb{ \bar{ X}}_R \Bbb{X}_R+ \beta \,  ( \Bbb{  X}_L \Bbb{ X}_R+c.c. ) \)\Big]\Big |\,.
}
Using the formulae above and integrating out the auxiliary superfields $\Psi_\pm$ gives
\equ{
\L_{matter}=\int \D_+ \D_- \,\( \frac{1}{2}\(g_{\mu \nu}+b_{\mu \nu}\)\D_+X^{\mu}\D_- X^{\nu}+2 i\, \sigma_I \mu_I\)\,,
\label{semichiral lagrangian appendix}
}
where $g,b$ are the flat-space metric and $b$-field of the ungauged case (set $a=b=1, d=\beta$ in \ref{g and b flat space}) and
\eqs{\nonumber
\mu_{1}\equiv \,& \bar X_{L}X_{L}+ \bar X_{R}X_{R}+\beta  (X_{L} X_{R}+\bar X_{L}\bar X_{R})\,,\\ \label{definition moment maps app}
\mu_{2}\equiv \,&-( \bar X_{L}X_{L}- \bar X_{R}X_{R})\,,\\ \nonumber
\mu_{3}\equiv \,&- \frac{i \beta}{2}\(X_{L}X_{R}-\bar X_{L}\bar X_{R}\)\,.
}
This is not the standard way of writing 
the matter $\N=(1,1)$ Lagrangian; in 
\cite{Hull:1989jk,Jack:1989ne,Kapustin:2006ic}, it is written in a way that makes clear the invariance of the action under $b$-field transformations $\delta b = d \lambda$ for an arbitrary 1-form $\lambda$. If we write the gauge transformations of the matter fields 
as $\delta X^\mu=\xi^\mu$ where $X^\mu =\{X_L,\bar X_L, X_R,\bar X_R\}$ and 
$\xi^\mu =iQ\{X_L,-\bar X_L, -X_R,\bar X_R\}$ for semichiral matter fields with charge $\pm Q$, then we can rewrite (\ref{semichiral lagrangian appendix})
as
\eqs{\nonumber
\L_{matter}=&~\L_1+\L_2\,,\\
\L_1=&\int D_+ D_- \,\( \frac{1}{2}g_{\mu \nu}\D_+X^{\mu}\D_- X^{\nu}+2 i\, \sigma_I \mu_I\)\,,\\
\L_2=&\int D_+D_-\,
\(\frac12\left[b_{\mu\nu}D_+X^\mu D_-X^\nu -
(\Gamma_+D_-X^\mu+\Gamma_-D_+X^\mu)(b_{\mu\nu}\xi^\nu)\right]\)\,.
\label{semichiral lagrangian2 appendix}
}
where $\D_\pm X^\mu=D_\pm X^\mu-\Gamma_\pm\xi^\mu$. The first term in (\ref{semichiral lagrangian2 appendix}) is just the ungauged one and behaves as expected under $b$-transformations. The second term can be rewritten as 
\equ{-\int D_+ D_- \,\frac12(\Gamma_+D_-X^\mu+\Gamma_-D_+X^\mu)
\, \alpha_\mu\,,}
where $\alpha_\mu$ is defined by $\iota_\xi (d b)\equiv d\alpha$, $\iota_\xi\alpha=0$, which is not changed by a $b$-transformation. In our case 
\equ{\alpha_\mu =\partial_\mu(\frac12b_{\rho\nu}X^\rho\xi^\nu)
\propto \partial_\mu \( 2iQ(-X_LX_R+\bar X_L \bar X_R)\)\,. }

\subsubsection*{\it Couplings to the CSVM}

In the case of the CSVM, we need to impose the constraint (\ref{Lphi 1,1}); this modifies the $\N=(1,1)$ action, leaving only one moment map coupling. As discussed in \cite{Kapustin:2006ic} the one-form $\alpha$ must satisfy 
\equ{
g J_{\pm} \xi\mp J^{t}_{\pm} \alpha = d \mu\,, \qquad \iota_\xi \alpha=0\,,
}
for the model to admit $\N=(2,2)$ supersymmetry. In our case these are satisfied automatically with $\alpha_\mu=b'_{\mu\nu}\xi^\nu$ and $\mu=4 \mu_2$.

\subsection{Reduction to $\N=0$ and the scalar potential}
\label{Reduction to N=0 and scalar potential}

In a NLSM, reduction to $\N=(1,1)$ is sufficient to study geometric data such as the metric, $b$-field, or complex structures. However, to compute the scalar potential in a GLSM one must reduce to $\N=0$ components. The reduction of a bosonic  $\N=(1,1)$ superfield to $\N=0$ components is given by\footnote{To avoid introducing new notation here we denote the $\N=(1,1)$ multiplet and its lowest component by the same symbol; it should be clear from the context what is meant.}
\eqsn{
X=\,&X\Big|\,, \qquad \psi_\pm= \D_\pm X\Big |\,, \qquad g=\D_+\D_- X\Big|\,,\\
\bar X=\,&\bar X\Big|\,, \qquad \bar \psi_\pm= \D_\pm \bar X\Big |\,, \qquad \bar g=\D_+\D_-\bar X\Big|\,.
}
Here $|$ means setting the remaining $\N=(1,1)$  Grassmann variables to zero. 

To compute the scalar potential, we further reduce  (\ref{semichiral lagrangian appendix}) to $\N=0$ components and integrate out the remaining auxiliary fields. We denote the lowest scalars of $\Bbb F$ and $\tilde{\Bbb F}$ by $\sigma$ and $\tilde \sigma$, respectively, the auxiliary scalars by $D_I=\D_+\D_- \sigma_I|$ and the auxiliary scalars in the semichiral multiplet by $g_{L,R}=\D_+\D_- X_{L,R}|$. Integrating out $g_{L,R}$ we obtain the relevant terms
\eqs{\nonumber 
\L=\,&-\frac{1}{2e^{2}}\(D_{1}^{2}+D_{2}^{2}+\frac{1}{2} D_{3}^{2} \)-\beta^{2}\(|\sigma|^{2}+\tfrac{1}{\beta^{2}-1}|\tilde \sigma|^{2}\)  \frac{1}{2}|X|^2\\ \nonumber
\,&+ 2 i \(\bar X_{L}\, D_{1} X_{L} -\bar X_{R} D_{1} X_{R}-\beta  X_{L}D_{1}X_{R}+\beta  \bar X_{L}D_{1}\bar X_{R} \)\\ \nonumber
\,& -  2i\,  \(\bar X_{L} D_{2} X_{L} +\bar X_{R} D_{2} X_{R}  \)\\
\,&-\beta (X_{L}D_{3}X_{R}+\bar X_{L}D_{3}\bar X_{R} )+...
\label{lagrangian scalar potential all Ds} 
}
where $|X|^2>0$ is contracted with the flat-space metric $g$ in (\ref{g and b flat space beta}) and the ellipses represent kinetic and fermionic terms that do not contribute to the scalar potential. 

For the unconstrained SVM one must integrate out all $D_I$, finally leading to the scalar potential
\equ{
U=2 e^{2} (\mu_{1}-r_{1})^{2}+ 2 e^2 (\mu_{2}-r_{2})^{2}+4 e^2 (\mu_{3}-r_{3})^{2}+\beta^{2}\(|\sigma|^{2}+\tfrac{1}{\beta^{2}-1}|\tilde \sigma|^{2}\)  \frac{1}{2}|X|^2\,,
}
where the functions $\mu_I$ are defined in (\ref{definition moment maps app}). 

For the constrained SVM with $\Bbb F=M$ one must set $D_1=D_3=0$ and $\sigma=M$ in (\ref{lagrangian scalar potential all Ds}). Integrating out $D_2$ gives 
\equ{
U= 2 e^2 (\mu_{2}-r_2)^{2}+\beta^{2}\(|M|^{2}+\tfrac{1}{\beta^{2}-1}|\tilde \sigma|^{2}\)  \frac{1}{2}|X|^2\,.
}
Similarly, the scalar potential for the constrained SVM with $\tilde{\Bbb F}=\tilde M$ is obtained by setting $D_2=D_3=0$ and $\tilde \sigma=\tilde M$ in (\ref{lagrangian scalar potential all Ds}), leading to the scalar potential
\equ{
\tilde U= 2 e^2 (\mu_{1}-r_{1})^{2}+\beta^{2}\(|\sigma|^{2}+\tfrac{1}{\beta^{2}-1}|\tilde M|^{2}\)  \frac{1}{2}|X|^2\,.
}
It is easy to see that the moduli space in the case $\Bbb F=M$ and $\tilde{\Bbb F}=\tilde M$ coincide. For example, the equation $\mu_1=r_1$ can be written as: 
\equ{
\frac{1}{2}(\beta+1)|X_L+\bar X_R|^2 - \frac{1}{2}(\beta-1)|X_L-\bar X_R|^2 = r_1\,.
}
Assuming $\beta>1$ the space of solutions to this equation coincides with the space of solutions to $\mu_2=r_2$.

For completeness, we give the potential in the case of equal charges and a CSVM with $\Bbb F=M$:
\eqsn{
U=2e^{2}(\mu_{2}'-r_2)^{2}+\alpha^{2}\(\tfrac{1}{(\alpha^{2}-1)}|M|^{2}+|\tilde \sigma|^{2}\)  \frac{1}{2}|X|^2  \,,
}
where $|X|^2$ is contracted with the corresponding flat-space metric and $\mu_2'= \bar X_L X_L+\bar X_R X_R+\alpha (\bar X_L X_R+c.c.)$. 

\subsection{Components of Semichiral multiplets}
\label{Components of Semichiral fields}

We denote the $\N=0$ components of a left semichiral multiplet by $(X_L,F_L,M_{-+},\psi^l_{\pm},\bar \chi_{-}^{l},\bar \eta_{-}^{l},M_{--})$ and their conjugates, and for right semichiral multiplet by $(X_R,F_R,M_{+-},\psi^r_{\pm},\bar \chi_{+}^{r},\bar \eta_{+}^{r},M_{++})$ and their conjugates. These are defined by  
\eqsc{X_{L}=\Bbb X_{L}\big|\,, \qquad \psi_{\pm}^{l}= \, \Bbb D_{\pm}\Bbb X_{L} \big|\,, \qquad   F_{L}= \Bbb D_{+}\Bbb D_{-}\Bbb X_{L}\big|\,, \nonumber\\ \label{components XL}
\bar{ X}_{L}=\bar{\Bbb X}_{L}\big|\,, \qquad \bar{\psi}^{l}_{\pm}= \,\bar{\Bbb D}_{\pm}\bar{\Bbb X}_{L} \big|\,, \qquad   \bar{ F}_{L}= \bar{\Bbb D}_{+} \bar{\Bbb D}_{-}\bar{\Bbb X}_{L}\big|\,, \\ 
\bar \chi_{-}^{l}= \bar{ \Bbb D}_{-}\Bbb X_{L}\big| \,, \quad  M_{-+}=\, \Bbb D_{+}\bar{\Bbb  D}_{-} \Bbb X_{L}\big| \,,\quad  M_{--}=\Bbb D_{-} \bar{ \Bbb D}_{-}\Bbb X_{L} \big|\,, \quad  \bar \eta_{-}^{l}= \Bbb D_{+} \Bbb D_{-}\bar{ \Bbb D}_{-} \Bbb X_{L}\big| \,,\nonumber\\ 
\chi_{-}^{l}= \Bbb D_{-}\bar{\Bbb X}_{L}\big| \,, \quad  \bar M_{-+}=\,\bar{\Bbb D}_{+} \Bbb  D_{-} \bar{\Bbb X}_{L}\big| \,,\quad  \bar M_{--}=\bar{\Bbb D}_{-}  \Bbb D_{-} \bar{\Bbb X}_{L} \big|\,, \quad  \eta_{-}^{l}=\bar{ \Bbb D}_{+} \bar{ \Bbb D}_{-} \Bbb D_{-} \bar{ \Bbb X}_{L}\big| \nonumber \,,
}
and 
\eqsc{X_{R}=\Bbb X_{R}\big|\,, \qquad \psi_{\pm}^{r}= \, \Bbb D_{\pm}\Bbb X_{R} \big|\,, \qquad   F_R= \Bbb D_{+}\Bbb D_{-}\Bbb X_{R}\big|\,, \nonumber\\ \label{components XR}
\bar{ X}_{R}=\bar{\Bbb X}_{R}\big|\,, \qquad \bar{\psi}^{r}_{\pm}= \,\bar{\Bbb D}_{\pm}\bar{\Bbb X}_{R} \big|\,, \qquad   \bar{ F}_{R}= \bar{\Bbb D}_{+} \bar{\Bbb D}_{-}\bar{\Bbb X}_{R}\big|\,, \\ 
\bar \chi_{+}^{r}= \bar{ \Bbb D}_{+}\Bbb X_{R}\big| \,, \quad  M_{+-}=\, \Bbb D_{-}\bar{\Bbb  D}_{+} \Bbb X_{R}\big| \,,\quad  M_{++}=\Bbb D_{+} \bar{ \Bbb D}_{+}\Bbb X_{R} \big|\,, \quad  \bar \eta_{+}^{r}= \Bbb D_{+} \Bbb D_{-}\bar{ \Bbb D}_{+} \Bbb X_{R}\big| \,,\nonumber\\ 
\chi_{+}^{r}= \Bbb D_{+}\bar{\Bbb X}_{R}\big| \,, \quad  \bar M_{+-}=\,\bar{\Bbb D}_{-} \Bbb  D_{+} \bar{\Bbb X}_{R}\big| \,,\quad  \bar M_{++}=\bar{\Bbb D}_{+}  \Bbb D_{+} \bar{\Bbb X}_{R} \big|\,, \quad  \eta_{+}^{r}=\bar{ \Bbb D}_{+} \bar{ \Bbb D}_{-} \Bbb D_{+} \bar{ \Bbb X}_{R}\big| \nonumber \,.
}
Setting $\chi=M=\eta=0$ in either of these multiplets reduces the field content to the off-shell content of a chiral multiplet. With appropriate identifications, it also contains the field content of the twisted chiral multiplet.

When the fields are coupled to a certain vector multiplet, one defines the `hatted' fields (as discussed in Appendix~\ref{Vector Multiplets}) and the component fields are defined by using gauge-covariant derivatives. 

\bibliographystyle{utphys}
\bibliography{References}

\end{document}